\documentclass{article}

\usepackage[top=1.7cm,bottom=1.2cm,left=1.4cm,right=1.4cm,includehead,includefoot]{geometry}
\usepackage{amsmath}
\usepackage{amsfonts}
\usepackage{amsthm}
\usepackage{graphicx}
\usepackage{subfigure}
\usepackage{color}
\newcommand{\dif}{\mathrm{d}}

\theoremstyle{plain}

  \theoremstyle{definition}
  
  \newtheorem{example}{Example}
  \theoremstyle{remark}
  \newtheorem{remark}{\textmd{Remark}}

\begin{document}

%%%%% title and author(s):
% \markboth{Author(s)}{Short Title}
% \title{Title}

%\markboth{X. Y. Kong, Z. F. Zhang and Z. D. Huang}{X. Y. Kong, Z. F. Zhang and Z. D. Huang}
\title{Piecewise constant level set algorithm for an inverse elliptic problem
in nonlinear electromagnetism}

%% single author:
%% \author[AUTHOR]{AUTHOR\corrauth}
%% \address{address of AUTHOR}
%% \email{{\tt email address of AUTHOR} (AUTHOR)}
%\author[Only Author]{Only Author\corrauth}
%\address{School of Mathematical Sciences, Beijing International University,
%Beijing 12345, China.}
%\email{{\tt email@address} (Only Author)}

% multiple authors:
% Please mark \corrauth after the name of the corresponding author.
% different addresses:
%\author[AUTHOR1 and AUTHOR2]{AUTHOR1\affil{1}\comma\corrauth and AUTHOR2\affil{2}}
%\address{\affilnum{1}\ address of AUTHOR1\\
%\affilnum{2}\ address of AUTHOR2}

\author{Xiangyin Kong\thanks{consosion@163.com}, Zhengfang Zhang\thanks{zhengfangzhang@hdu.edu.cn} and Zhengda Huang\thanks{zdhuang@zju.edu.cn}\\{\footnotesize \emph{Department of Mathematics, Zhejiang University, Hangzhou, P.R. China, 310027}}\\{\footnotesize \emph{Department of Mathematics, Hangzhou Dianzi University, Hangzhou, Zhejiang, 310018, P. R. China}}\\{\footnotesize \emph{Department of Mathematics, Zhejiang University, Hangzhou, P.R. China, 310027}}}

%
%same address:
%%\author[AUTHOR1, AUTHOR2 and AUTHOR3]{AUTHOR1, AUTHOR2\corrauth and AUTHOR3}
%\address{\affilnum{1}\ Department of Mathematics, Zhejiang University, Hangzhou, P.R. China, 310027 \\
%          \affilnum{2}\ Department of Mathematics, Hangzhou Dianzi University, Hangzhou, Zhejiang, 310018, P. R. China}
%%
%\emails{{\tt consosion@163.com}(X.\;Y.\;Kong),\;{\tt zhengfangzhang@hdu.edu.cn}(Z.\;F.\;Zhang),\\~~\;{\tt zdhuang@zju.edu.cn}(Z.\;D.\;Huang)}
%
\maketitle
%%%%% Begin Abstract %%%%%%%%%%%
\begin{abstract}
An inverse problem of identifying inhomogeneity or crack in the workpiece made of nonlinear magnetic material is
 investigated. To recover the shape from the local measurements, a piecewise constant level set algorithm is proposed.  By means of the Lagrangian multiplier method, we derive the first  variation w.r.t the  level set function  and obtain the descent direction   by the adjoint variable method. Numerical results show the robustness and effectiveness of our algorithm applied to reconstruct some complex shapes.
\end{abstract}
%%%%% end %%%%%%%%%%%

%%%%% Keywords %%%%%%%%%%%

%%%% AMS subject classifications %%%%
%\ams{65K10, 35Q60, 74M10}

%%%% maketitle %%%%%
%\maketitle

%%%% Start %%%%%%
\section{Introduction}\label{section1}

%%%%%%%%%%%%%------the content of Section 1-----
In many applications, one needs to find out the flaws in materials nondestructively. Inspired by this, the non-destructive evaluation technique has attracted the eyes of many researchers during the last decade. As one kind of non-destructive evaluation technique, eddy current testing technique \cite{ECT} has been used for flaw detection. In this paper, we intend to design an algorithm to identify the crack or inhomogeneities in the nonlinear magnetic material like  steel from the local measurements of the magnetic induction.

This inverse problem is a problem of shape reconstruction. Similar to general shape recovery problem, there is no information about the interface of the optimal shape as a prior, so we need to have a good mechanism to express the shape and track the evolution of the shape. The level set method was first originally proposed by Osher and Sethian in \cite{osher1}. For this method, the interface between two adjacent domains is represented by the zero level set of a Lipschitz continuous function. Through the change of this function, this method can easily handle many types of shape and topological changes, such as merging, splitting and developing sharp corners. Due to these merits, it has been used in various areas, such as epitaxial growth \cite{epixialgrowth}, inverse problem,  optimal design \cite{optimaldesign}, image segmentation \cite{imageseg},  structure topology optimization \cite{structureopti1} and EIT problem \cite{EIT1}. For the sake of the numerical stability, the level set function is usually chosen as the  signed distance function, but at most cases, the level set function after each iteration is not be a signed distance function and is  usually re-initialized by solving an ordinary differential equation \cite{reinitialization}. In \cite{IC_ref}, Cimr\'{a}k et al. used the level set method to represent the shape of the inhomogeneity and evolve the shape by minimizing a functional during the iterative process. In \cite{IC_thermal,IC_sinum}, Cimr\'{a}k also used the level set method for the representation of the interface to solve some inverse problems in thermal imaging and the nonlinear ferromagnetic material.  As to the initial value of the level set function, it was reported in \cite{levelweakness1,levelweakness2,topoder1} that the level set method based only on the shape sensitivity may get stuck at shapes with fewer holes than the optimal geometry in some applications such as structure designs.
When one wants to use the level set method to solve the practical problem, he can reduce the effects of the initial value  of the level set function on the final results in the next two ways.  The first way is to choose the shape with
enough holes as the initial value. The second one is to introduce the topological derivative  into the level set method to let the shape  create holes in the iterations \cite{topoder1,topoder2,topoder3,topoder4,topoder5}. To find a good  initial value of the level set function, the researchers in \cite{IC_ref} proposed the gradient-for-initial approach which is based on the idea that the domain which can drop the value of the cost functional should be the air gap. In that approach, the parameter in smeared-out Heaviside function should be set large enough. The large value of this parameter, however,  may cause the oscillation phenomenon. So the parameter in the acquisition of the initial choice of the level set function and the evolving  process of level set function should be set separately.

%
%\mathcal{F}

Recently, piecewise constant level set method, which is a variant of level set method, was proposed by Lie, Lysaker and Tai in \cite{pclproposal1,pclproposal2,pclproposal3}.  To distinguish these two methods, we call the former one traditional level set method. Unlike the traditional level set method, the interface between two adjacent sub-domains is represented by the discontinuity of a piecewise constant level set function. Compared with traditional level set method, piecewise constant level set method has at least two advantages. One merit is that it can create many small holes automatically without the topological derivatives during the iterative process. Furthermore, it is verified by many numerical examples that the final result is independent of the initial value of the level set function in many numerical tests. And the other one is that the piecewise constant level set method  need not to re-initialize the level set function periodically during the evolution process, thus,  reduces the computational cost a lot. Since it was proposed, it has been applied in various fields  such as image segmentation, elliptic inverse coefficient identification, optimal shape design, electrical impedance tomography and positron emission tomography \cite{Heavidesmooth}. Lie, Lysaker and Tai took this method to solve the image segmentation in
\cite{ pclproposal1,pclproposal2,pclproposal3,Tai3}  and the elliptic inverse problem and interface motion problem in
\cite{pclforInverse,pclforinterface}. Wei and Wang used piecewise constant level set method to solve structural topology optimization in \cite{pclforsto}. Zhu, Liu and Wu applied the piecewise constant level set method to solve a class of two-phase shape optimization problems in \cite{Zhu1}. Zhang and Cheng proposed a boundary piecewise constant level set method to deal with the boundary control of eigenvalue optimization problems in \cite{Zhang1}. In this paper, we intend to use the piecewise constant level set method to represent the shape and recover the exact shape of the crack or the inhomogeneities of nonlinear magnetic material.

%%%%%%%%%%%%%%%%%%%%%%%%%%%%%%%%%%%%%%%%%%%%%

Based on the piecewise constant level set method, we propose a piecewise constant level set algorithm to recover the shape of the nonlinear magnetic material problem. We introduce the piecewise constant level set method and convert the constrained optimization problem into an unconstrained one by  the Lagrangian multiplier method. In the numerical tests, our algorithm relies little on the initial guess of the level set function.  Moreover, our algorithm can reconstruct shape accurately even when the noise level is high.

The rest of this paper is organized as follows. In Section 2, we introduce the piecewise constant level method in brief. In Section 3, we  introduce the direct problem and the inverse problem. In Section 4, we describe the deduction of the first variation w.r.t level set function for the objective functional in the unconstrained optimization problem in detail. In Section 5, we  present numerical results to show the effectiveness and robustness of piecewise constant level set algorithm.

\section{Piecewise constant level set method}\label{section2}

We first introduce the piecewise level set method in brief. Suppose that  the domain $\Omega$ is the union of some  sub-domains
$\Omega_i,\ i=1,2,\cdots,m$, i.e.,  \begin{equation*}
  \bar{\Omega}=\bigcup\limits_{i=1}^{m}\Omega_i\cup \bigcup\limits_{i=1}^{m}\partial\Omega_i,
  \end{equation*} where $\partial\Omega_i$ is the boundary of  sub-domain $\Omega_i$.

  If there exists a function $\phi$  defined as
  \begin{equation}
  \label{pclfunction}
  \phi(\emph{\textbf{x}})=i, \quad  \emph{\textbf{x}}\in \Omega_i,\ \ \text{where}\ \  i=1,2,\cdots,m,
  \end{equation}
 then the interface between two adjoint sub-domains can be identified by the discontinuity of the function $\phi$ and the characteristic function of the $i$-th sub-domain $\Omega_i$ can be written as
  \begin{equation}
  \label{characterfunc}
  \chi_i=\prod\limits_{j=1,j\neq i}^{m} \frac{(\phi-j)}{(i-j)}.
  \end{equation}
%With  the formulation of  $\chi_i$ in (\ref{characterfunc}), the length of the boundary of
%the domain $\Omega_i$ can be obtained by $$|\partial
%\Omega_i|=\int_{\Omega}|\nabla\chi_i|\dif \emph{\textbf{x}}.$$

For the traditional level set method proposed in \cite{osher1}, if
 $\phi$ is a level set function, then the expression
of $\chi_i$  contains the Heaviside function $H(x)$, which is defined
as below:
\begin{equation}\label{Hfunc}H(x)=\left\{\begin{array}{ll}1,&x\geq
0,\\0,&x<0.
\end{array}\right.\end{equation}  Since $H(x)$ is not differentiable at 0, it
is often replaced by  smooth functions which contain some
parameters in \cite{pclforinterface}. Sometimes the parameters in the replaced
smooth functions can cause difficulties in the result analysis,
such as the convergence of the level set function $\phi$ in \cite{IC_ref}.
For the piecewise constant level set method, we do not encounter this
kind of problems.

If $\phi$ is defined as in (\ref{pclfunction}), then it holds
that
\begin{equation}
   \label{Kvalue}
   K(\phi)=0, \quad \text{in} \; \Omega,
   \end{equation}
where the function $K$ is defined as
 \begin{equation}
   \label{Kfunc}
   K(\phi)=(\phi-1)(\phi-2)\cdots(\phi-m)=\prod\limits_{i=1}^m(\phi-i).
  \end{equation}

   It should be pointed out that if  Eq. (\ref{Kvalue}) holds, then every point $\textbf{\emph{x}}\in \Omega$
is in  one and only one sub-domain.  In other words,  there is
no vacuum and overlap between two different sub-domains.

%
%   %% need to add some materials........

%
   For any piecewise smooth function $f(\textbf{\emph{x}})$ that  coincides with  $f_i(\textbf{\emph{x}})$  in $\Omega_i$,  it can be written  as
   \begin{equation}
   f(\emph{\textbf{x}})=\sum\limits_{i=1}^mf_i(\emph{\textbf{x}})\chi_i(\phi(\textbf{\emph{x}})),
   \end{equation}
   where $\phi(\textbf{\emph{x}})$ is defined in (\ref{pclfunction}).

   %%%%%%%%%%%%%%%%%%%%%%%%%%%%%%% New   Section %%%%%%%%%%%%%%%%%%%%%%%%%%%%%%%%%%%%%%%%%%%%%%%%%%%%%%%%%%%%%%
\section{Direct problem and inverse problem}\label{section3}

We first introduce the quasi-linear  partial differential equation which will be used  in this paper.

%Since we need to solve an direct problem, we first introduce the
%direct problem.

This equation  is  defined as
\begin{equation}
\label{modelpde}\left\{
\begin{array}{ll}
\nabla\cdot(v(\emph{\textbf{x}},|\nabla A|^2)\nabla A)=J,&
\text{in}\; \ \:\Omega,
\\\quad A=0\,, &\text{on}\; \partial\Omega,
\end{array}
\right.
\end{equation}
 where $\Omega\subset \mathbb{R}^2$ is a bounded domain with $C^{1}$
boundary,  the symbol $J$ denotes a suitable function defined in
$\Omega$ and the  function $v:\Omega\times \mathbb{R}\rightarrow
\mathbb{R}$ is   defined as
\begin{equation}
\label{vfunction} v(\emph{\textbf{x}},s)=\left\{
\begin{array}{ll}
v_1(s),&\qquad \emph{\textbf{x}}\,\in D,\\
v_2(s),&\quad \emph{\textbf{x}}\,\in \Omega\backslash D,
\end{array}\right.
\end{equation} where $D\subset\Omega$, $v_1,v_2$ are two functions determined by the specific practical applications.

Eq. (\ref{modelpde}) can model many industrial and physical applications. When both $v_1$ and $v_2$ are constant functions,
it is the elliptic inverse problem in \cite{pclforInverse}. When $v_1$ and
$v_2$ are both linear ones, it  characterizes the electric
impendence tomograph problem \cite{EITpde}. Since the material in this paper is nonlinear magnetic, we use  the nonlinear model, which  is more robust and able to reconstruct the shape of the crack or inhomogeneities even when the linear model is not \cite{IC_ref}. In this model,  $v_2$ is a nonlinear function.

Since we need to solve the direct problem in each iteration, we next present
 the direct problem in detail. The direct problem is to obtain the quantity $A$ when the variable
$v$, the domain $D$ and the function $J$ are given. In this paper,
our goal is to design a numerical scheme to identify the crack or
the inhomogeneity in the workpiece which is made of nonlinear magnetic material. In this case, the function $J$ denotes the induced
current density, the quantity $A$ denotes the only nonzero component
of the vector potential \emph{\textbf{A}} which is perpendicular to
the $xy-$plane. According to the physical knowledge, the magnetic
induction $\mathbf{B}$ is
\begin{equation}
 \label{Bexpre}
 \mathbf{B}=(\frac{\partial A}{\partial y}, -\frac{\partial A}{\partial x},0)^T.
 \end{equation}
Since $A$ is defined on a subregion of $\mathbb{R}^2$,
we obtain
\begin{equation*}
 |\mathbf{B}|=|\nabla A|.
 \end{equation*}

After introducing two symbols $\mu_1$ and $\mu_2$ to denote the
magnetic permeability of the air and the nonlinear magnetic
material, we define the
functions $v_1$ and $v_2$ as the reciprocal of the two functions
$\mu_1$ and $\mu_2$, respectively,  i.e. $v_1=1/\mu_1$ and $\
v_2=1/\mu_2$. As to $\mu_2$, due to the nonlinear magnetic property of the material, it  depends on $|\nabla A|$. Usually, the
function $\mu_2(s)$ is monotonically increasing on the interval $[0, s_{max}]$ and
is monotonically decreasing on the interval $[s_{max}, +\infty]$, where $s_{max}$ is a number determined by the property of the magnetic nonlinear
material.
%%%%%%%%%%%%%%%%%%%%%%%%%%%%%%%%%%%%%%%%%%%%%%%%%%%%%%%%%%%%%%%%%%%%%%%%%%%%%%%%%%%%%%%%%

With the piecewise constant level set method,  the function $v(\emph{\textbf{x}})$ can be formulated as \begin{equation}
\label{vformula} v(\emph{\textbf{x}})=v_1(2-\phi)+v_2(\phi-1).
\end{equation}
By (\ref{vformula}), the weak form of Eq. (\ref{modelpde}) can be written as
\begin{equation}
\label{weakform} \int_{\Omega}[v_1(|\nabla A|^2)(2-\phi)\!+\!v_2(|\nabla
A|^2)(\phi-1)] \!\nabla A\! \cdot \!\nabla\varphi\dif
\emph{\textbf{x}}\!=\!\int_{\Omega}J \varphi\dif \emph{\textbf{x}},\ \forall \varphi\in \!W_{0}^{1,2}(\Omega).
\end{equation}
 The existence and uniqueness of the solution to the Eq. (\ref{weakform}), have been proved in \cite{Gilbarg,Ladyzhenskaya}. To ensure the existence and uniqueness of the solution to Eq. (\ref{weakform}), we adopted the assumptions in \cite{IC_ref} which are  listed as below:
 \begin{itemize}\setlength{\itemsep}{0pt}
\item[\textbf{A1}] The function $v_i$ is non-decreasing;
\item[\textbf{A2}]$\lim_{s\rightarrow 0}v_i(s)=v_{min}>0$;
\item[\textbf{A3}] $\lim_{s\rightarrow \infty}v_i(s)=v_{max}>0$ and define $v(s)=v_{max}$ for $s=\infty$;
\item[\textbf{A4}] $v_i$ is differentiable with well-defined derivatives $v_i^{'}$ satisfying
$$v_{min}^{'}\leq v_i^{'}\leq v_{max}^{'},$$
\end{itemize}
where $i=1,2$.

In order to let the function $\mu_2$ satisfy the non-decreasing property in the assumption \textbf{A1}, we set the function  $J$ large enough. As to the solution to Eq. (\ref{modelpde}), we consider it as the solution to the following nonlinear operator equation $$G(A)=J,$$ where $G$ is an operator defined in the space $W_{0}^{1,2}(\Omega)$ and takes values in $W_{0}^{1,2}(\Omega)$. To solve this nonlinear operator equation, we first choose an initial guess $A_0$ of $A$,
 and use the Newton-Raphson algorithm to update $A$, i.e.,
 $$A_{i+1}=A_{i}-[DG(A_i)]^{-1}(G(A_i)-J),$$
 where $i=1,2,\cdots$.

Now we introduce the inverse problem. When $v_1$ and $v_2$ are
given, the inverse problem is to reconstruct the actual shape of $D$
from the  measurement data $\overline{\mathbf{M}}$ of magnetic
induction in a given subset $\Gamma\subset\Omega$. Under the
framework of the piecewise constant level set method, the problem is
to find a piecewise constant function $\phi$ to  approximate the
exact level set function  which can represent the actual composition
of the workpiece (during the iterative process, the level set function $\phi$ may be not a piecewise constant one). In order to distinguish these two functions, we
 denote by $\phi_{exact}$ the exact level set function  here and
afterward. For our work, $\Gamma=\Omega$.

%In this paper, we study the workpiece made of the hard steel and
%possibly contain the crack which is filled with the air. The
%magnetic permeability of the air is close to 1, i.e. $\mu_1=1$, so
%$v_1=1$. For $v_2$, it is defined as
%$$v_2(s)=d_1+\frac{c_1s^{b_1}}{a_1^{b_1}+s^{b_1}},$$
%where the concrete values for the four variables in  $v_2$ are set
%the same as in \cite{IC_ref}:
%$$a_1=0.5,\quad b_1=4,\quad c_1=3,\quad d_1=0.2.$$
%%%%%%%%%%%%%%%%%%%%%%%%%%%%%%%%%%%%%%%%%%%%%%%%%%%%%%%%%%%%%%%%%%%%%%%%%%%%%%%5
When the level set function $\phi$ and  the data
$\overline{\mathbf{M}}$ on $\Gamma$ are given, we use the following
functional
\begin{equation} \label{Misfit}
F_1(\phi)=\frac{1}{2}\int_{\Gamma}|\nabla
A(\phi)-\overline{\mathbf{M}}|^2 \dif \emph{\textbf{x}},
\end{equation} to measure the misfit between  $\phi$ and
$\phi_{exact}$, where
$\overline{\mathbf{M}}=(-\overline{B}_2,\overline{B}_1)^T$ with
$\overline{B}_1$ and $\overline{B}_2$ being the x-axis and y-axis
component of $\mathbf{B}$. Obviously, the smaller the value of $F_1$
is, the more close the level set $\phi$ is to $\phi_{exact}$. Thus
the inverse problem is converted into the following optimization
problem
\begin{equation}\label{constrained}
\min\limits_{\phi}F_1(\phi)\quad \text{subject to } K(\phi)=0,
\end{equation}
where $K(\phi)$ is defined in (\ref{Kfunc}).

\section{Algorithm}\label{section4}
%%%%%%%%%%%%%%%%%%%%%%%%%%%%%%% New Section   %%%%%%%%%%%%%%%%%%%%%%%%%%%%%%%%%%%%%%%%%%%%%%%%%%%%%%%%%%%%%%

In this section, we will introduce our piecewise constant level set algorithm for this inverse
problem.

We first give the  deduction of the gradient of $F_1$. Introducing two symbols $DF_1$ and $\delta_{h}g$ to denote the
gradient of $F_1$ and the G\^{a}teaux derivative of $g$ with respect to $\phi$
in the direction $h$, we give the deduction of $\delta_{h}F_1$.
According to the definition,  it can be expressed as
\begin{equation}
\label{Fgradient} \delta_{h}F_1=\lim_{\epsilon\rightarrow
0}\frac{F_1(\phi +\epsilon
h)-F_1(\phi)}{\epsilon}=\int_{\Gamma}\nabla\delta_{h}A\cdot(\nabla
A-\overline{\mathbf{M}})\dif \emph{\textbf{x}},
\end{equation}
 where  $\nabla\delta_{h}A$ will be given below .

 Differentiating both sides of Eq. (\ref{weakform}) with respect to $\phi$, we obtain the following
sensitivity equation
 \begin{equation}
 \label{derfor1}
\int_{\Omega}\delta_{h}[v(\mathbf{x},|\nabla A|^2)]\nabla A\cdot
\nabla\varphi\dif
\emph{\textbf{x}}+\int_{\Omega}v(\emph{\textbf{x}},|\nabla
A|^2)\nabla\delta_{h}A\cdot \nabla\varphi\dif \emph{\textbf{x}}=0.
 \end{equation}

Since $A$ is completely determined by the level set function $\phi$ when
$J$, $v_1$ and $v_2$ are given,   from (\ref{vformula}), the integral
part of the first term in (\ref{derfor1}) can be sorted as
 \begin{equation}\label{deltav}
 \begin{aligned}
 \delta_{h}[v(\emph{\textbf{x}},|\nabla A|^2)]  = &(v_2(|\nabla A|^2)-v_1(|\nabla A|^2))h\\&+(2-\phi)\delta_{h}[v_1(|\nabla A|^2)]+(\phi-1)\delta_{h}[v_2(|\nabla A|^2)]\\
 =&(v_2(|\nabla A|^2)-v_1(|\nabla A|^2))h\\&+2((2-\phi)v_1^{'}(|\nabla A|^2)+(\phi-1)v_2^{'}(|\nabla A|^2))\nabla A\cdot\nabla\delta_{h}A.
 \end{aligned}
 \end{equation}
Substitute (\ref{deltav}) into (\ref{derfor1}) and simplify the
equation,  the sensitivity equation (\ref{derfor1}) can be written
as
 \begin{equation}
 \label{sensieq}
 \begin{aligned}
 &\int_{\Omega}(v_2(|\nabla A|^2)-v_1(|\nabla A|^2))h\nabla A\cdot \nabla\varphi\dif \emph{\textbf{x}}\\&=
 -\int_{\Omega}2((2-\phi)v_1^{'}(|\nabla A|^2)+(\phi-1)v_2^{'}(|\nabla A|^2))\nabla A\cdot\nabla\delta_{h}A \nabla A\cdot \nabla\varphi \dif \emph{\textbf{x}}\\
 &\quad-\int_{\Omega}v(\emph{\textbf{x}},|\nabla A|^2)\nabla\delta_{h}A\cdot \nabla\varphi\dif \emph{\textbf{x}}.
 \end{aligned}
 \end{equation}
  As to the computation of $\nabla\delta_{h}A$, it could be deduced by the adjoint variable
  method which was used in many problems \cite{AdjointMethod1,AdjointMethod2,AdjointMethod3,AdjointMethod4,AdjointMethod5,AdjointMethod6}.
 Specifically, we take the following steps to avoid the direct computation of $\nabla\delta_{h}A$.

 Firstly, we determine the solution  to the following equation which we will
  denote $p$ afterward
  \begin{equation}
  \label{adjointequation}
  \begin{aligned}
\int_{\Omega}2((2-\phi)v_1^{'}(|\nabla A|^2)+(\phi-1)v_2^{'}(|\nabla A|^2))&\nabla p\cdot\nabla A \nabla A\cdot \nabla\psi \dif \emph{\textbf{x}}
\\+\int_{\Omega}v(\mathbf{x},|\nabla A|^2)\nabla p\cdot \nabla\psi\dif \emph{\textbf{x}}&=\int_{\Gamma}\nabla\delta_{h}A\cdot(\nabla
A-\overline{\mathbf{M}})\dif \emph{\textbf{x}}.
  \end{aligned}
  \end{equation}
Secondly, by letting the  test function for the variable $\varphi$ and
$\psi$ in (\ref{sensieq}) and (\ref{adjointequation}) being $p$ and
$\delta_{h} A$, respectively, we  can get the following equation
\begin{equation}
\label{compareeq}
\begin{aligned}
-\int_{\Omega}(v_2(|\nabla A|^2)-v_1(|\nabla A|^2))h\nabla A\cdot
\nabla\psi\dif
\emph{\textbf{x}}=\int_{\Gamma}\nabla\delta_{h}A\cdot(\nabla
A-\overline{\mathbf{M}})\dif \emph{\textbf{x}}.
\end{aligned}
\end{equation}
 By comparing  Eq. (\ref{Fgradient}) with Eq. (\ref{compareeq}), we obtain
 \begin{equation}
 \delta_{h}F_1=-\int_{\Omega}(v_2(|\nabla A|^2)-v_1(|\nabla A|^2))h\nabla A\cdot \nabla p \dif \emph{\textbf{x}}.
 \end{equation}

 After deriving the formula of $\delta_{h}F_1$, we simply obtain $DF_1$ by simply projecting $(v_2(|\nabla A|^2)-v_1(|\nabla A|^2))\nabla A\cdot \nabla p$ onto the finite element space $W_{0}^{1,2}(\Omega)$ by solving the following equation
 \begin{equation}
 \label{DF_1eq}
 \begin{aligned}
 -\int_{\Omega}(v_2(|\nabla A|^2)-v_1(|\nabla A|^2))\nabla A\cdot \nabla p \varphi \dif \emph{\textbf{x}}=\int_{\Omega}DF_1\varphi \dif \emph{\textbf{x}}, \  \forall \varphi\in W_{0}^{1,2}(\Omega).
 \end{aligned}
 \end{equation}

In the numerical tests, it often comes out that the level set
function $\phi$ can not approximate $\phi_{exact}$ correctly at the boundary of $\Omega$. To approximate the function $\phi_{exact}$ more accurately, we
introduce a Tikhonov stabilizing term $\int_{\Omega}|\nabla\phi|^2\dif \emph{\textbf{x}}$, then the constrained optimization problem
(\ref{constrained}) becomes
\begin{equation}
\label{constrained_regur} \min\limits_{\phi}F,\quad \text{subject to
} K(\phi)=0,
\end{equation}
where $F=F_1+\alpha\int_{\Omega}|\nabla \phi|^2\dif \emph{\textbf{x}}$ and the
coefficient $\alpha$ is a positive real number.

Similar to the deduction of $DF_1$, the gradient $DF$ of $F$ is calculated
 by solving the following equation
\begin{equation}
 \label{DFeq}
 \begin{aligned}
 \int_{\Omega}(v_2(|\!\nabla A\!|^2)\!-\!v_1(|\nabla A|^2))\nabla A\cdot \nabla p \varphi \dif \emph{\textbf{x}}\!+\!2\alpha \!\int_{\Omega}\nabla\phi\cdot \nabla p \varphi \dif \emph{\textbf{x}} \!=\!\int_{\Omega}DF\varphi \dif \emph{\textbf{x}},\forall \varphi\in\! W_{0}^{1,2}(\Omega).
 \end{aligned}
 \end{equation}

In this paper, we use
the Lagrangian multiplier method to convert the problem (\ref{constrained_regur}) into the following
unconstrained optimization problem
\begin{equation}
\label{unconstrained} L(\phi)=F(\phi)+\int_{\Omega}l_1(x)K(\phi)\dif
\emph{\textbf{x}},
\end{equation}
where $l_1$ is the Lagrangian multiplier, a $l^2$-integrable
function defined on the domain $\Omega$.

According to the general theory of optimization, the
level set function $\phi$ that we seek is the saddle
point of the functional $L(\phi)$, that is,
\begin{equation}
\label{saddle}
\begin{aligned}
\frac{\partial L}{\partial \phi}&=\frac{\partial F}{\partial \phi}+l_1(2\phi-3)=0, \\
 \frac{\partial L}{\partial l_1}&=K(\phi)=0.
 \end{aligned}
\end{equation}

By multiplying two sides of Eq. (\ref{Kvalue}) by
($2\phi-3$) and making use of the constraint $(\phi-1)(\phi-2)=0$,
we get the formula to  update the multiplier
\begin{equation}
\label{muformula} l_1=-(2\phi-3)\frac{\partial F}{\partial \phi}.
\end{equation}
Substituting (\ref{muformula}) into (\ref{saddle}), we have
\begin{equation}
\label{Lder} \frac{\partial L}{\partial
\phi}=-4(\phi-1)(\phi-2)\frac{\partial F}{\partial \phi}.
\end{equation}

We introduce the artificial time variable $t$ and update the level set function
$\phi$ according to  the following scheme
\begin{equation}
\label{levelupdateway} \left\{
\begin{array}{l}
\frac{\partial\phi }{\partial t}=-\frac{\partial L}{\partial
\phi}\qquad \text{in}\quad \Omega\times\mathbf{R}^{+},
\\\phi(\emph{\textbf{x}},t)=\phi_{0}(\emph{\textbf{x}}) \quad \text{in}\quad \Omega,
\end{array}\right.
\end{equation}
until the level set function $\phi$ satisfy $\frac{\partial
\phi}{\partial t}=0$.

To discretize (\ref{levelupdateway}), we use the forward Euler scheme
\begin{equation}
\label{levelupdatescheme}
\phi_{k+1}=\phi_{k}-\Delta t_k\frac{\partial L}{\partial \phi}\vert_{\phi=\phi_k}, \quad k=0,1,2,\cdots.
\end{equation}
During the iteration process, for the sake of numerical stability, we let the
time step $\Delta t_k$ satisfy the Courant-Friedrichs-Lewy
condition
  \begin{equation}
  \label{timeinterval}
  \Delta t_k=\sigma h/\max\limits_{\emph{\textbf{x}}\in \Omega}\vert\frac{\partial L}{\partial \phi}(\phi_k)\vert,
  \end{equation} where $\sigma\in (0,1)$ and $h$ is mesh size.

We now give the choice of $\phi_{0}$ and the
projection used in the iteration process. We  let $\phi_0$ be a  function whose value at every point is neither 1 nor 2, or let $\phi_0$ be a constant function and the constant is a number between 1 and 2 (1 and 2 are excluded).
Considering the fact that the value of the final $\phi$ at every point $\textbf{x}\in D$ should be either 1 or 2, we  project $\phi$ in the following
way:
\begin{equation}
  \label{projection}
  P_{\{1,2\}}(\phi)=\left\{
  \begin{array}{lr}
  1,& \quad \phi<1,\\
  2,& \quad \phi>2,\\
  \phi,&\quad \text{otherwise},
  \end{array}
  \right.
  \end{equation}
  after updating $\phi$ by (\ref{levelupdatescheme}) at each step.

  From Eq. (\ref{Lder}), the gradient $\frac{\partial L}{\partial \phi}$ is equal to 0 when the value of $\phi$ is either 1 or 2. This  sometimes causes the iterative
process unable to start or proceed, thus we exclude $1$ and $2$
from being the candidates of the constant for  $\phi_0$.
In order to avoid this phenomenon in the iteration process, we
count the number $N$ of the points at which the function $\phi$
take 1 or 2 after projecting the level set function $\phi$ by $P_{1,2}$. If the
number $N$ is equal to $N_T$ which denotes the number of the
total points, we stop and exit the iterative process.

In our numerical tests, there is no such case that the
iteration process stops for the reason that  $N=N_T$. Considering the fact that  $F_1$ becomes smaller as the iteration proceeds, we  introduce a variable $osci$ to denote the times that the $F_1$
oscillates and exit the iterative process when $osci$ reaches a given
number.

\noindent Now we will present the piecewise constant level set algorithm (PCLSA).

\textbf{Algorithm 1 PCLSA}\\
\indent Initialize $\phi_0$ as a suitable function, $F_{1,-1}=10000$
, $osci=0$ and $N=0$, Compute $N_T$. For $k=0,1,2,\cdots$,
\begin{description}\itemsep=-2pt
\item[Step 1.]  Use $\phi_k$ to update $v:v_k(\emph{\textbf{x}})=v_{1,k}(2-\phi_k)+v_{2,k}(\phi_k-1)$ and obtain $A_k$ and $\nabla A_{k}$
by solving Eq. (\ref{weakform}) with the Newton-Raphson  method.
\item[Step 2.] Compute $F_{1,k}$ by (\ref{Misfit}). let $osci:=osci+1$ if \texttt{$F_{1,k}>F_{1,k-1}$}.
If $osci$ is equal to the predetermined value, exit the iterative
process; Otherwise, go to Step 3.
\item[Step 3.]  Solve the equation (\ref{DFeq}) to
compute the gradient of $F_1$ according to $\phi_k$ and $\nabla A_k$,
and update $\frac{\partial L}{\partial \phi}\vert_{\phi_k}$.
\item[Step 4.] Set $\Delta t_{k}$ by (\ref{timeinterval}) and use the scheme
(\ref{levelupdatescheme}) to update the level set function $\phi$.
\item[Step 5.] Project $\phi$ by (\ref{projection}). Check the projected $\phi$ and obtain $N$. Exit the iterative process if $N=N_T$. Or else, set $k:=k+1$ and $\phi_k=\phi$, go to Step 1.
\end{description}
%\begin{remark}
%The number  of iterations $Iter$ needed by  Algorithm 1 has something with
%the value of $\sigma$. The larger the value of $\sigma$ is, the
%smaller $Iter$ is. But there are some cases that we choose small
%value for $\sigma$ to get an more satisfactory result.
%\end{remark}
\begin{remark}
After exiting
the iterative process, the value of the level set function $\phi$ at every point
 $\emph{\textbf{x}}\in\Omega$ usually doesn't satisfy the constraint
 $(\phi-1)(\phi-2)=0$. To make $\phi$ satisfy the constraint, we project $\phi$
 as below:
\begin{equation}
  \label{projectionfinal}
  P_{u}(\phi)=\left\{
  \begin{array}{lr}
  1,& \quad \phi\leq 1.5,\\
  2,& \quad \phi>1.5,\\
  \end{array}
  \right.
  \end{equation}
 after exiting the iteration process.
\end{remark}

\section{Numerical results}\label{section5}

In this section, four examples are solved by Algorithm 1 (PCLSA). The workpiece we studied in the numerical tests is made of the hard steel and possibly contains the crack which is filled with the air. The magnetic permeability of  air is close to 1, i.e. $\mu_1 = 1$, so $v_1 = 1$. For the function $v_2$, which is relevant to the magnetic permeability of the hard steel, is defined as
$$v_2(s)=d_1+\frac{c_1s^{b_1}}{a_1^{b_1}+s^{b_1}},$$
where the concrete values for the four variables in $v_2$ are set the same as \cite{IC_ref}:
$$a_1 = 0.5,\  b_1 = 4,\  c_1 = 3,\  d_1 = 0.2.$$

For all examples,  the domain $\Omega=[-0.5, 0.5]\times[-0.5,0.5]$ and all the numerical tests are
run on the PC with Intel Core 2 Duo 2.10 GHz processor and 2
GB RAM by the software Matlab 2010b. When solving the equation
(\ref{weakform}) and (\ref{DFeq}),  we divide the domain $\Omega$
  into some rectangles with  the size $h_x=h_y=h=1/dim$, where $dim$ is the number of rectangles in the
$x-$direction. For the measurement data
$\overline{\mathbf{M}}$, we generate it in the following steps: we first find a level set
function $\phi_{exact}$ to represent the shape of $D$
accurately, then solve the equation (\ref{weakform}) with the
Newton-Raphson algorithm to obtain the solution $A_{exact}$, finally
assign $\nabla A_{exact}$ to $\overline{\mathbf{M}}$. In order to
see the closeness of the function $\phi$ to
$\phi_{exact}$, we call the built-in `contour' command in Matlab to
plot the interface between two sub-domains after the test. For the
command `contour', we set the parameter for the number of the
contour line as 1 and choose the red-solid-line and the blue-dotted-line to denote the interface between two sub-domains
represented by $\phi_{exact}$ and $\phi$, respectively. In the
figures that depict the evolution of the level set function, the red part and the blue part represent the hard steel and the
air, respectively. In the
figures that depict the evolution of the level set function, the red part and the blue part represent the hard steel and the
air, respectively. In some
figures, the shapes of some graphics don't look like shapes they should be. For instance, the shape of the circle in Example 1 is more like an octagon. In
order to measure the misfit between the computed $\phi$ and
$\phi_{exact}$ precisely, we count the number of the points
where two functions take different values.

In this section, we do two groups of numerical tests to show the
effectiveness and robustness of PCLSA. The first group of
tests are based on the measurement data $\overline{\mathbf{M}}$
without noise. In order to show the flexibility of the choice of the initial guess of $\phi$, we set the function $\phi$ as some different initial values
and observe the final shapes. The second group of tests are based on the measurements
data $\overline{\mathbf{M}}$ with  a certain level of noise. As to the robustness of  PCLSA, our algorithm can reconstruct the shape precisely when the noise level is up to 15\%, which is superior to the algorithm in \cite{IC_ref}.
%%%%%%%%%%%%%%%%%%     Example-1      %%%%%%%%%%%%%%%%%
\begin{example}\label{example1}\ \ \ In this example, We use the same configuration as  in \cite{IC_ref}. The exact shape of the domain $D$ is a circle with the
center $(0.2,0.15)$ and the radius 0.1. And  the induced current
density $J$ is defined as follows:
 \begin{equation}\label{Jfunction}
 J(x,y)=\left\{\begin{array}{rr}
 J_1,& y>0.4,\\
 -J_1,& y<-0.4,\\
0,& \text{otherwise}
 \end{array}\right.
 \end{equation}
 where $J_1=500$. This choice of $J$  describes the
 case that the workpiece is wrapped  by the wires.
\end{example}

\begin{figure}
\centering \subfigure[The interfaces represented by $\phi_{exact}$
and $\phi$.]{
\label{example1:a} %% label for first subfigure
\includegraphics[width=2.7in]{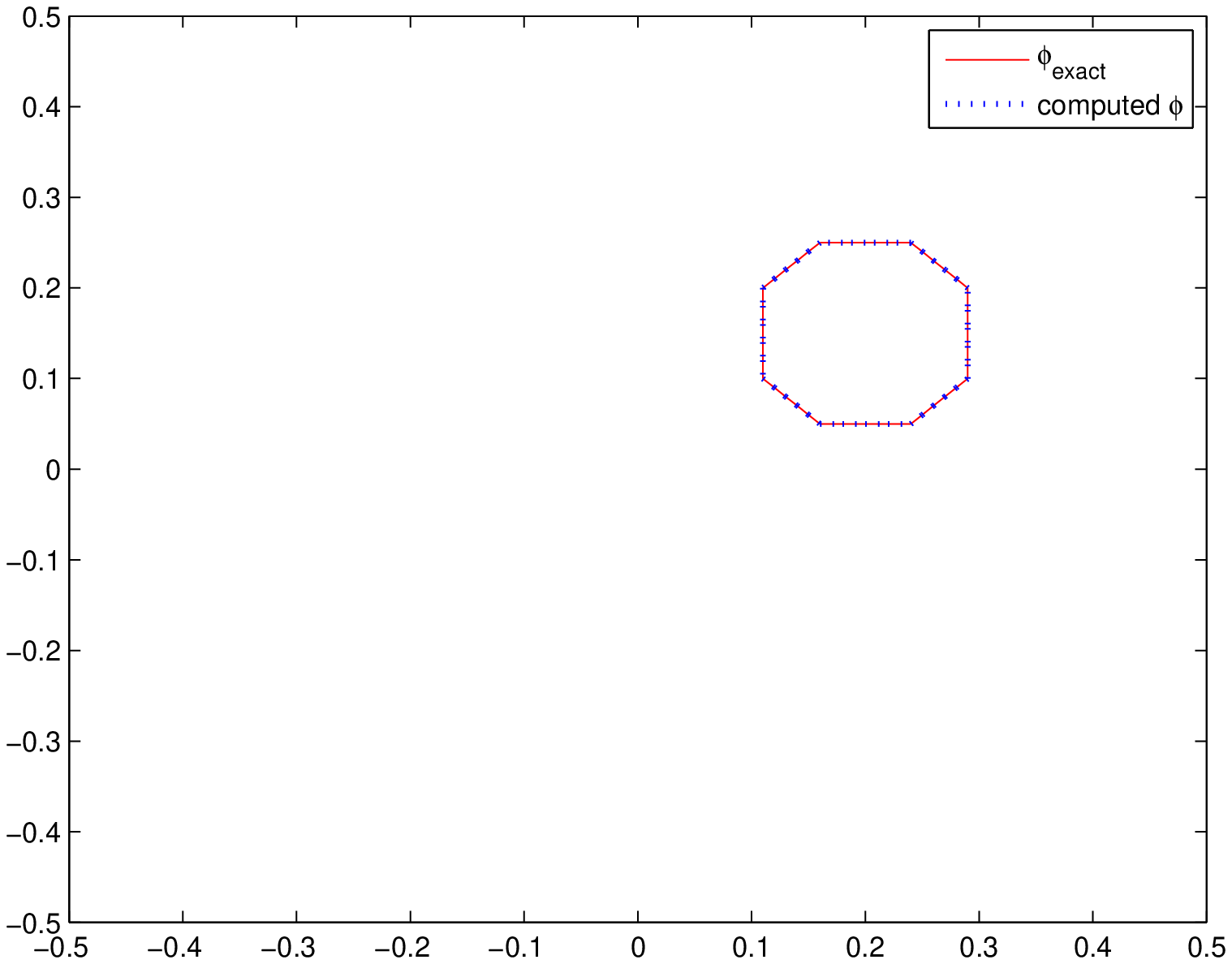}}
\hspace{0.1in} \subfigure[The value of $\log F_1/\log 10$.]{
\label{example1:b} %% label for second subfigure\\
\includegraphics[width=2.7in]{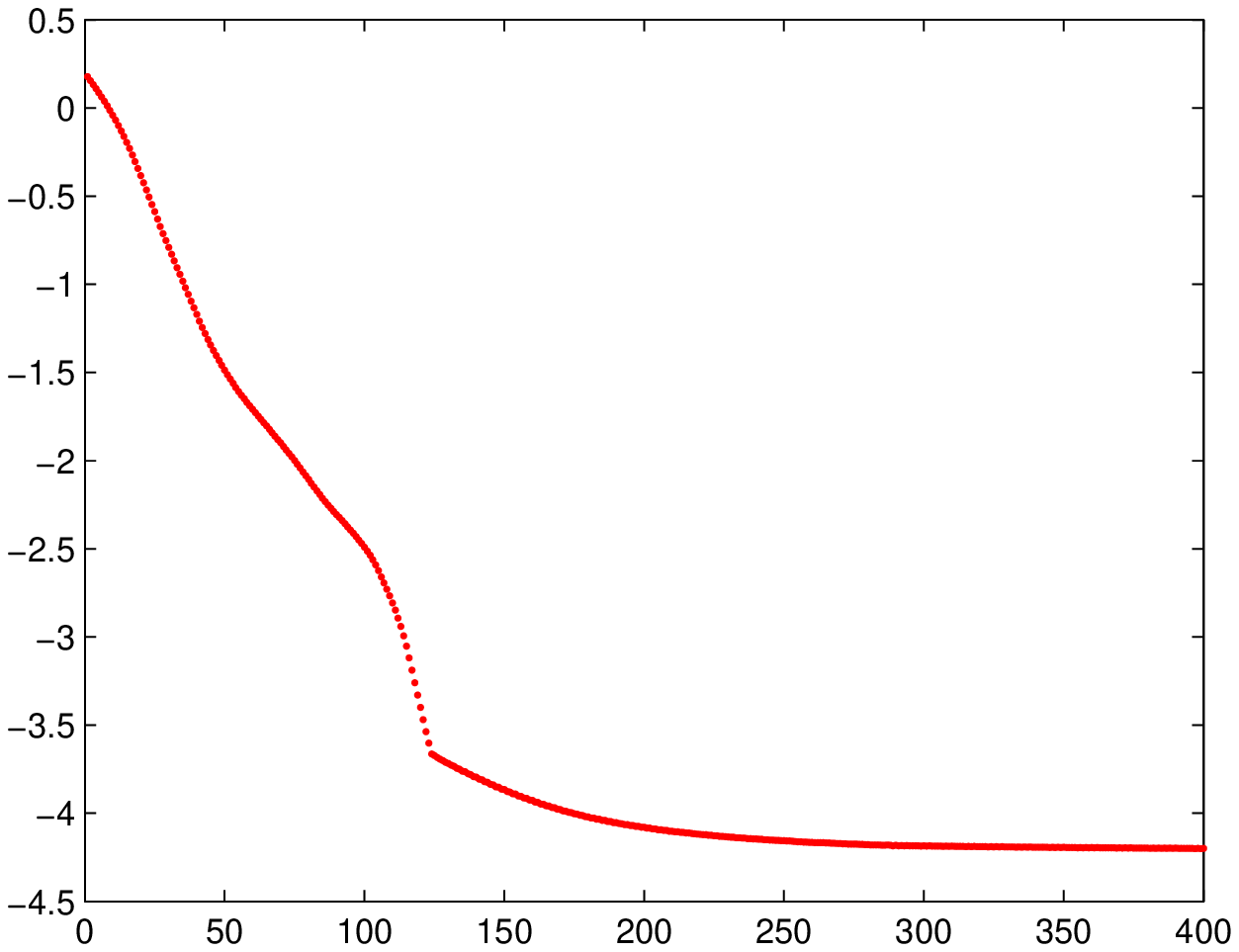}}
\hspace{0.3in} \subfigure[Step 75]{
\label{example1:c} %% label for third subfigure
\includegraphics[width=3.8in]{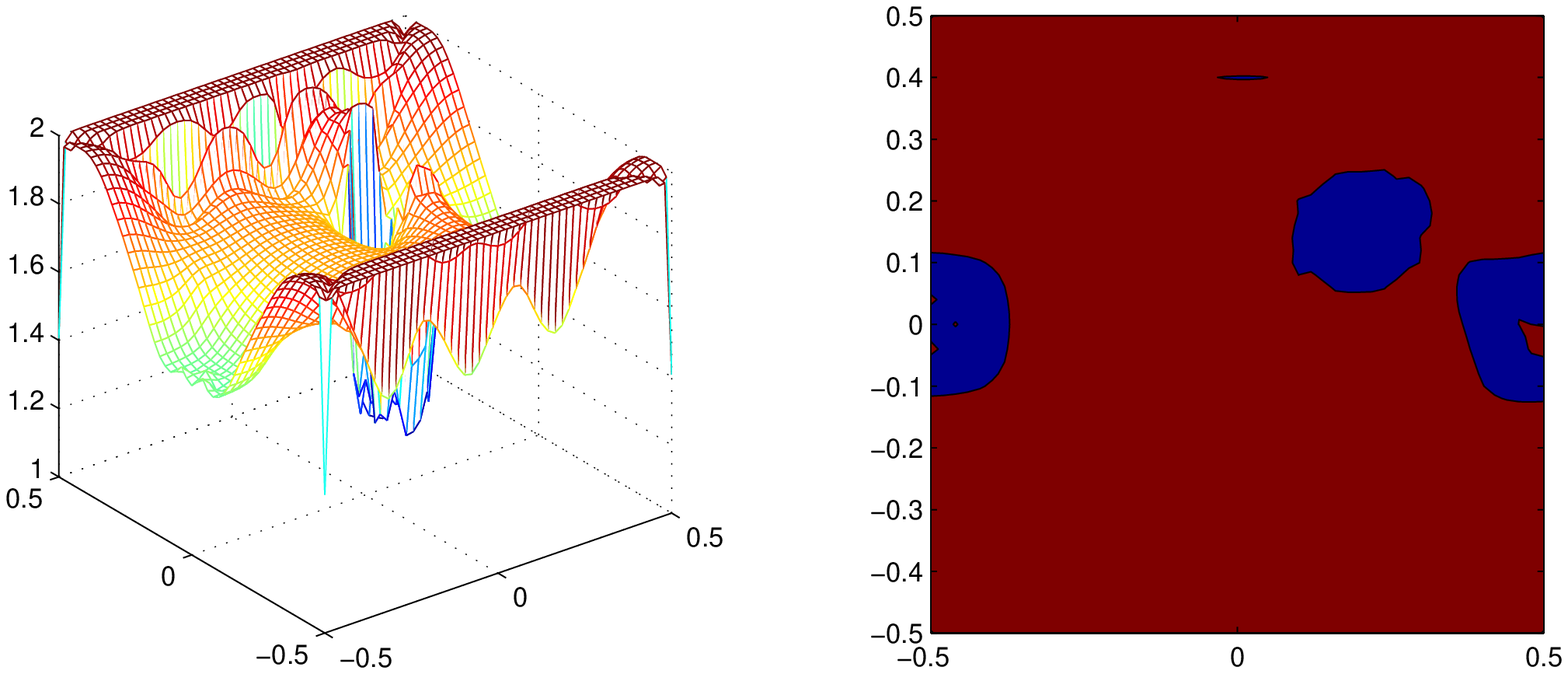}}
\hspace{0.3in} \subfigure[Step 150]{
\label{example1:d} %% label for fourth subfigure
\includegraphics[width=3.8in]{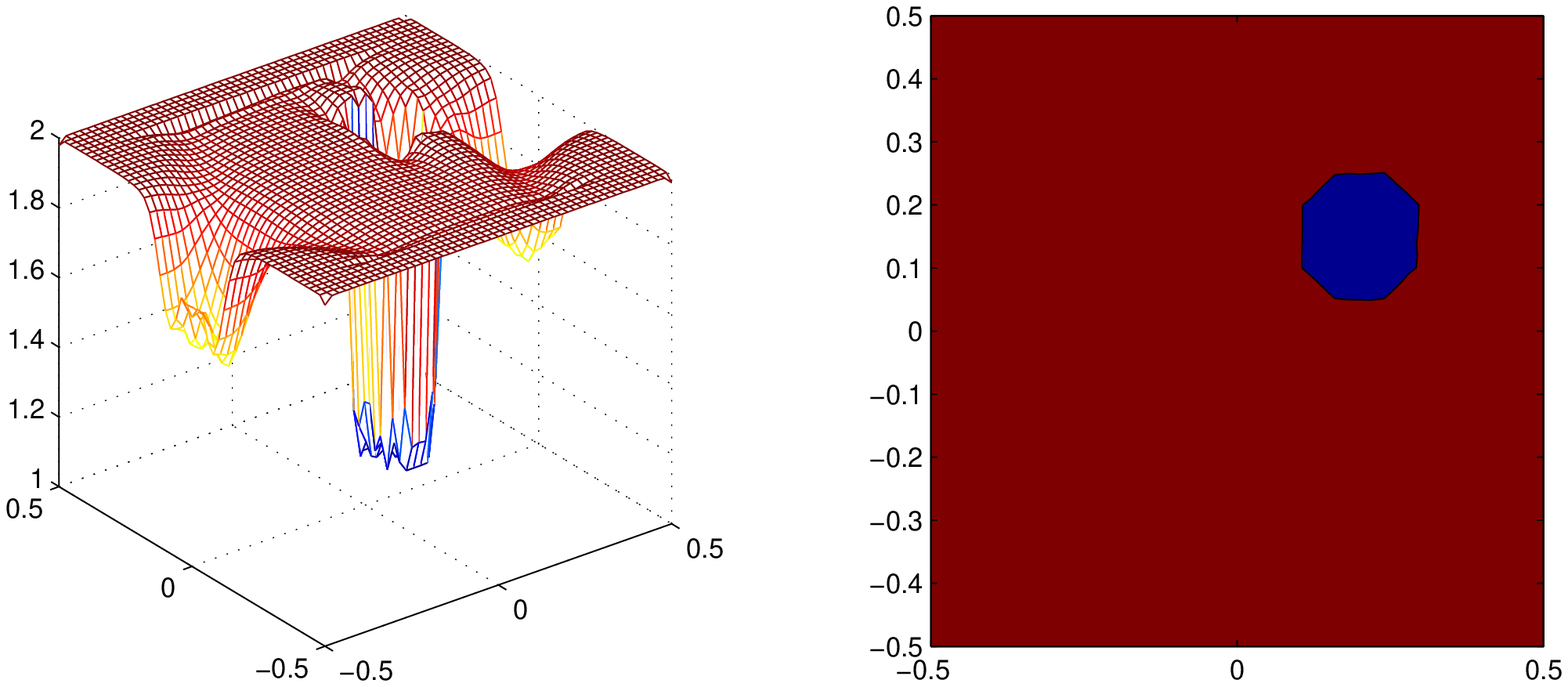}}
\hspace{0.3in} \subfigure[Step 225]{
\label{example1:e} %% label for fifth subfigure
\includegraphics[width=3.8in]{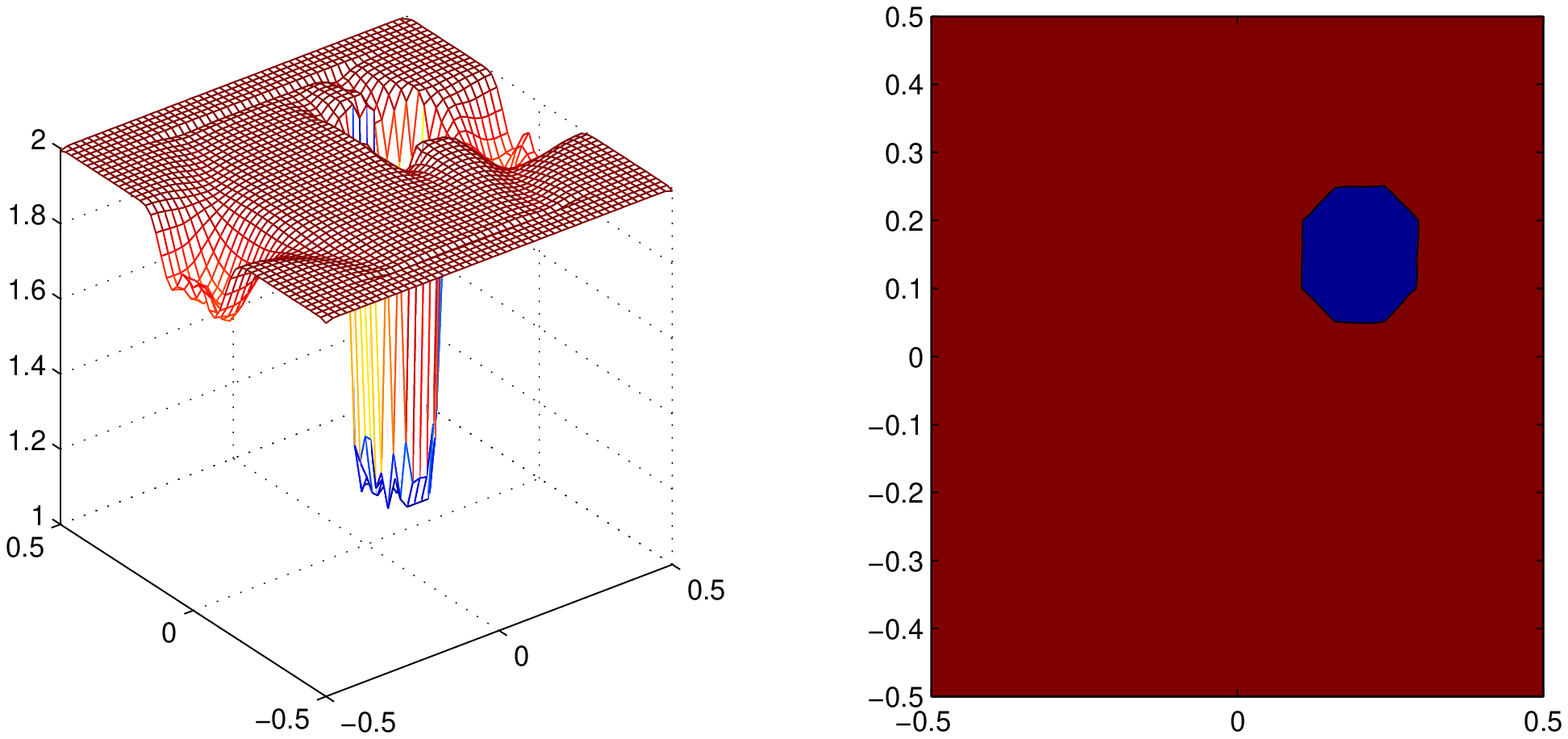}}
\hspace{0.3in} \subfigure[Step 300]{
\label{example1:f} %% label for sixth subfigure
\includegraphics[width=3.8in]{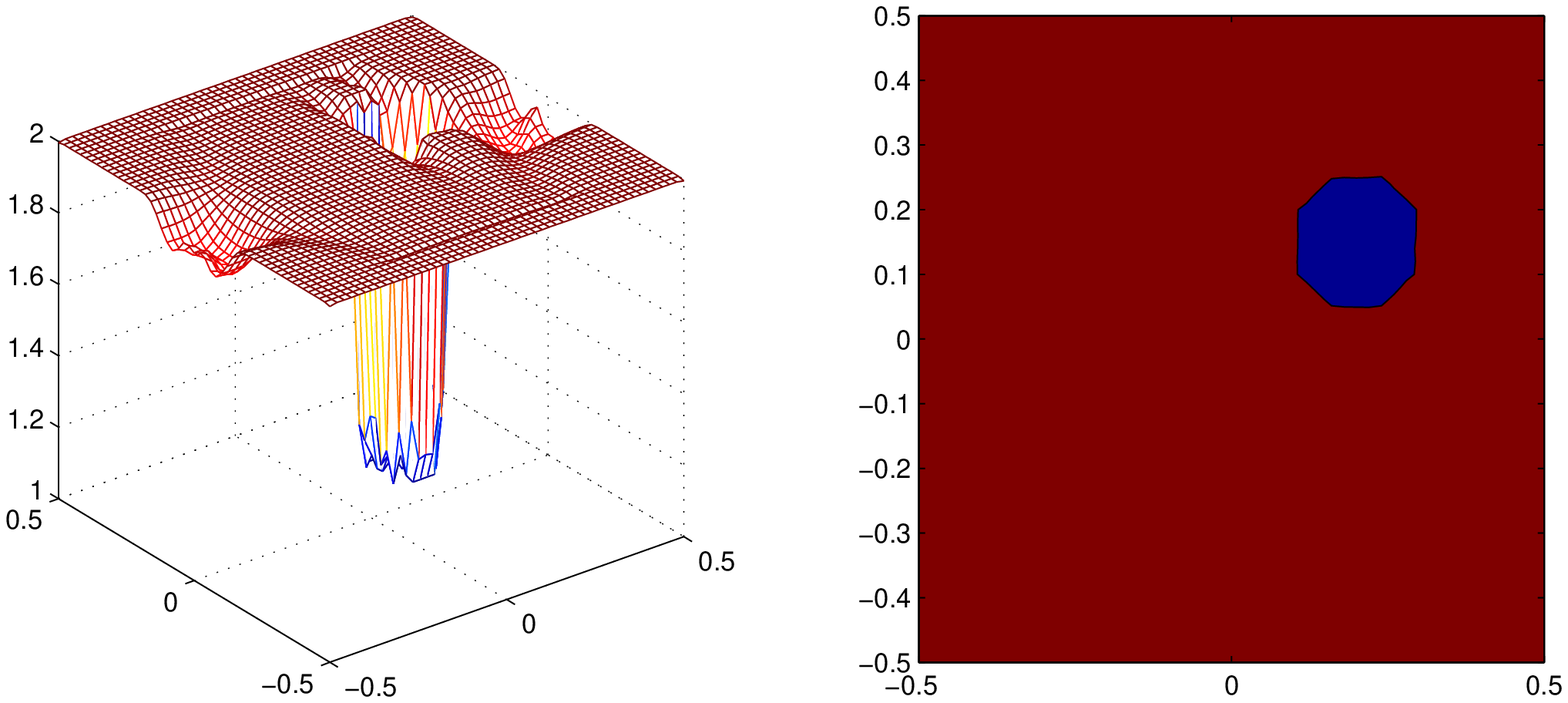}}
\caption{Numerical results for Example 1. (a) The final
reconstruction of $\phi$. (b) The change of the value of $F_1$ vs.
the number of iterations. (c)-(f) The evolution of $\phi$ and the interface.}
\label{example1:subfig} %% label for entire figure
\end{figure}

For this  test, we set $dim=50,\  \sigma=0.9\
\text{and} \ \alpha=0.001.$
 The initial value of the level set function is chosen as $\phi_{0}=1.5$ and the upper bound of  $osci$ is 10. The
iterative process stops after 304 steps. We present the results in Fig. \ref{example1:subfig}.

\begin{figure}
\centering
  \includegraphics[width=3in]{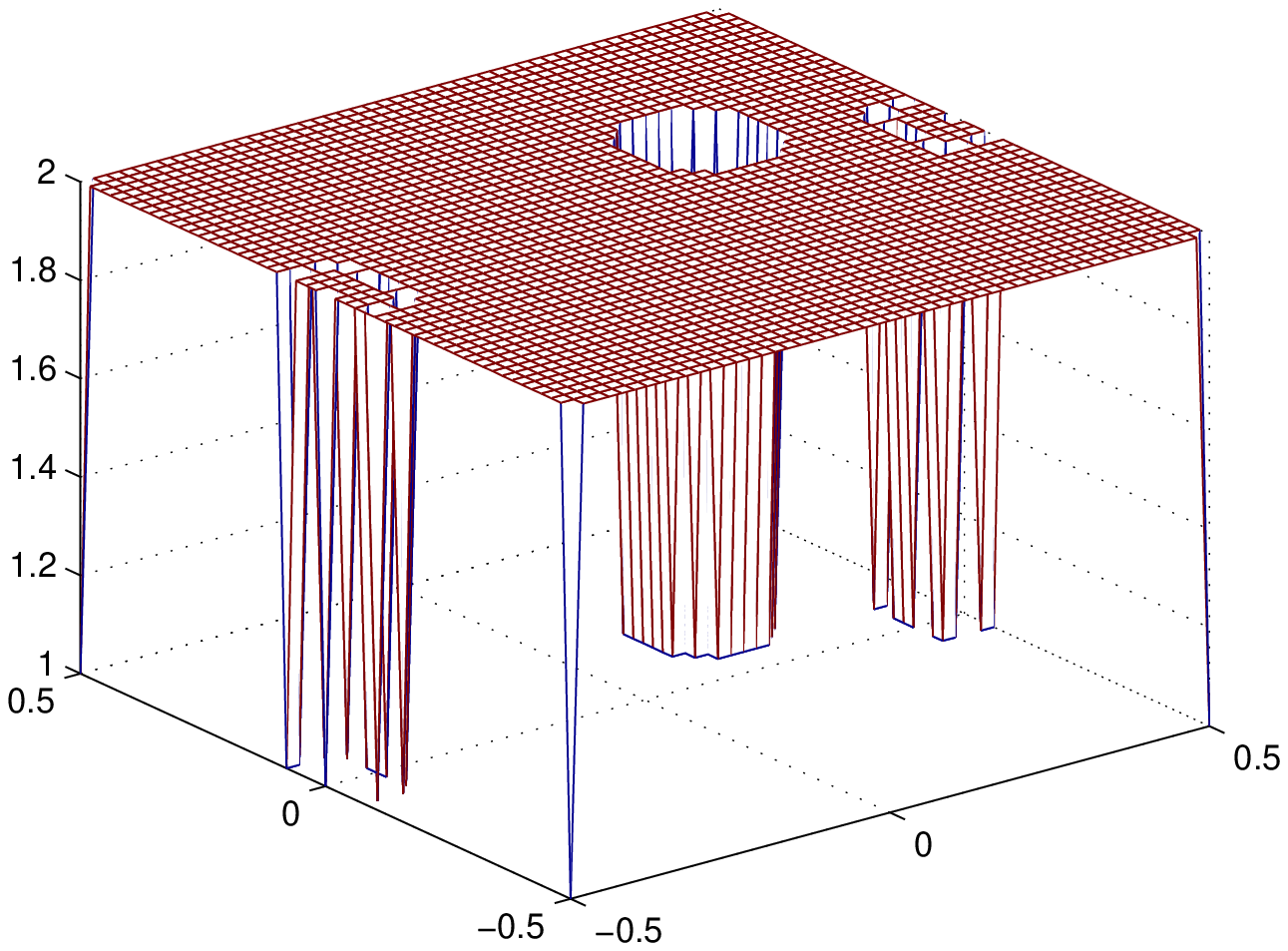}
  \caption{The figure of computed $\phi$  without regularization. }\label{example_compare_1}
\end{figure}

In Fig. \ref{example1:a}, the red solid line and the blue dotted line coincide, that is to say,  PCLSA can identify the shape
of the air gap completely.  In this test,
the two functions, $\phi$ and $\phi_{exact}$, take the same value at
every point $\textbf{\emph{x}}\in\Omega$. Fig. \ref{example1:b}
shows the change of $F_1$ with respect to the number of iterations.
From Fig. \ref{example1:b}, we can see that the value of $F_1$ first
becomes smaller, but doesn't become smaller any more after 300 steps.
So it is reasonable for us to choose $osci=10$. Fig.
 \ref{example1:c} to Fig. \ref{example1:f} show the evolution of the
 function $\phi$ and the interface. During the iterative process, the values of $\phi$  at the points of $\Omega$ develop towards to our expectation except
 four corner points. As the increase of the number of
iterations, the values of $\phi$ at these four points first become
small, which is away from our expectation, but becomes large after some steps. In Fig.
\ref{example_compare_1}, the values of the level set function $\phi$
 at the four corner points and some points at the boundary of
$\Omega$ are 1 instead of 2. By comparing with the picture of $\phi$
after 300 steps (see Fig. \ref{example1:f} and Fig.
\ref{example_compare_1}), we can see that it's the effects of the
regularization to let the values of $\phi$ at four points become
large.

%%%%%%%%%%%%%%%%%%     Example-2      %%%%%%%%%%%%%%%%%
\begin{example}\label{example2}
In this example, the shape of the air gap in the workpiece
 is  non-convex, which is more like a shrimp. The current density function $J$ is
defined as (\ref{Jfunction}) with the constant $J_1$  equal
to 500. This numerical test describes the case that the workpiece is
all wrapped with the wires.
\end{example}

\begin{figure}
\centering \subfigure[The interfaces represented by $\phi_{exact}$ and $\phi$.]{
\label{example2:a} %% label for first subfigure
\includegraphics[width=2.7in]{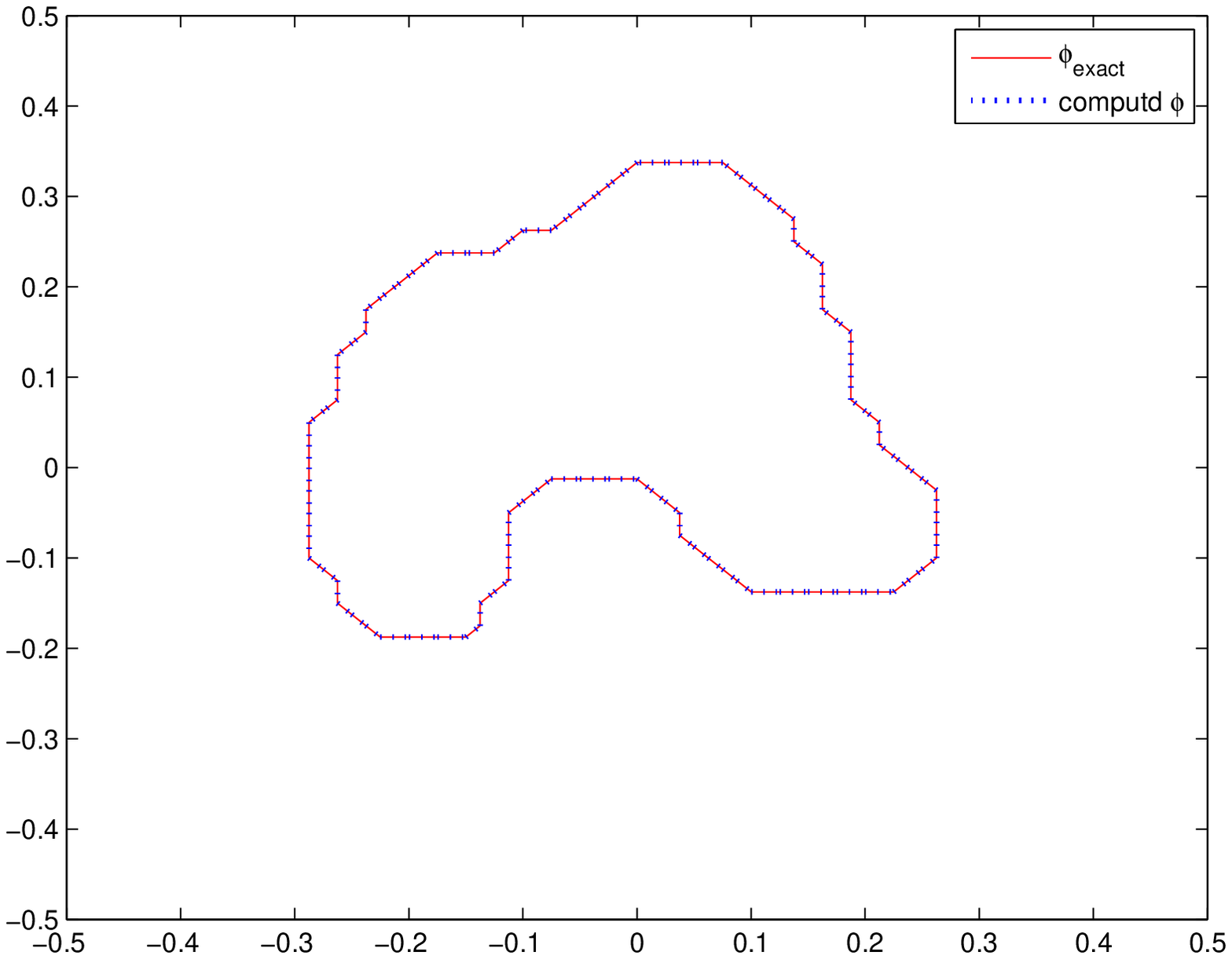}}
\hspace{0.1in} \subfigure[The value of $\log F_1/\log 10$.]{
\label{example2:b} %% label for second subfigure
\includegraphics[width=2.7in]{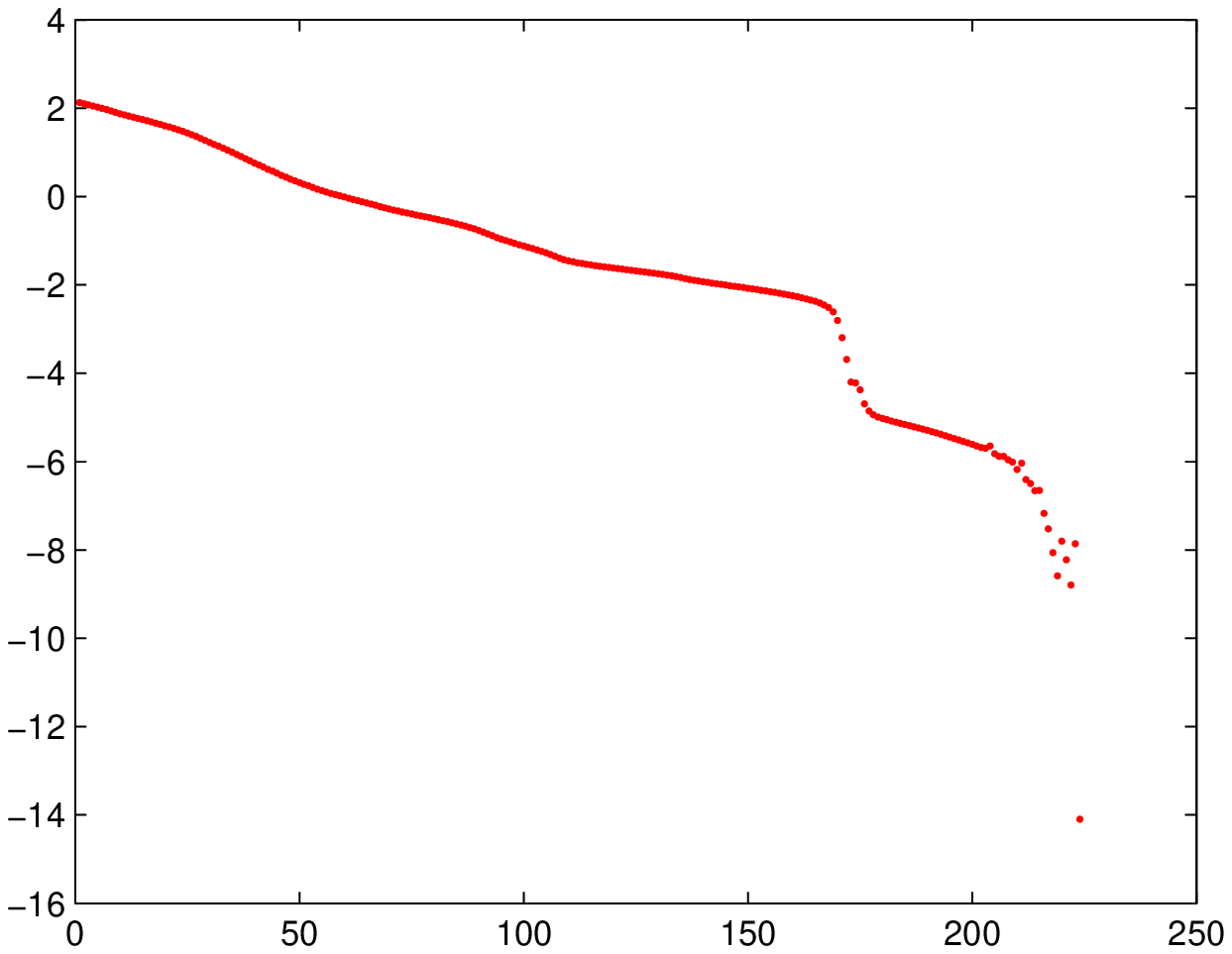}}\\
 \hspace{0.3in}\subfigure[Step 50]{
\label{example2:c} %% label for third subfigure
\includegraphics[width=3.8in]{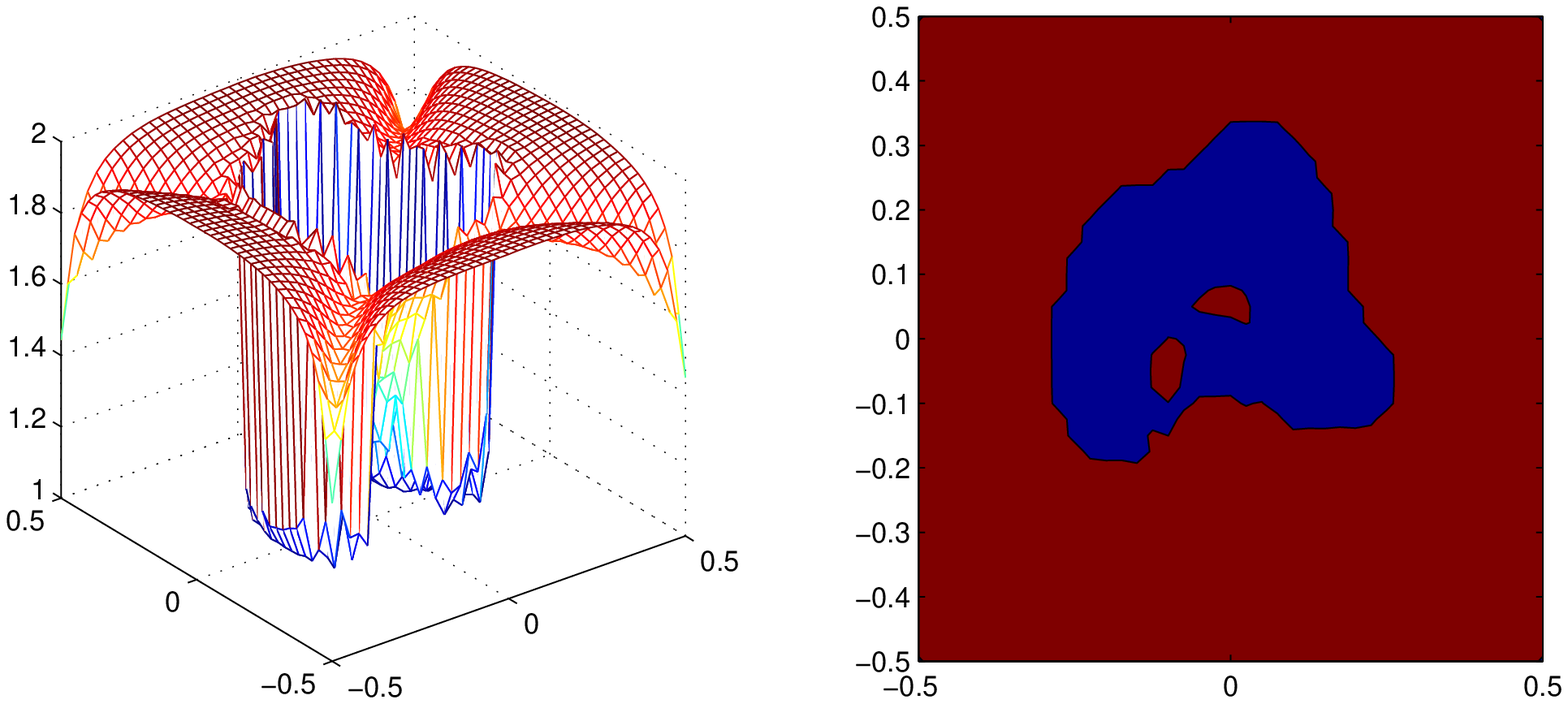}}
\hspace{0.3in} \subfigure[Step 100]{
\label{example2:d} %% label for fourth subfigure
\includegraphics[width=3.8in]{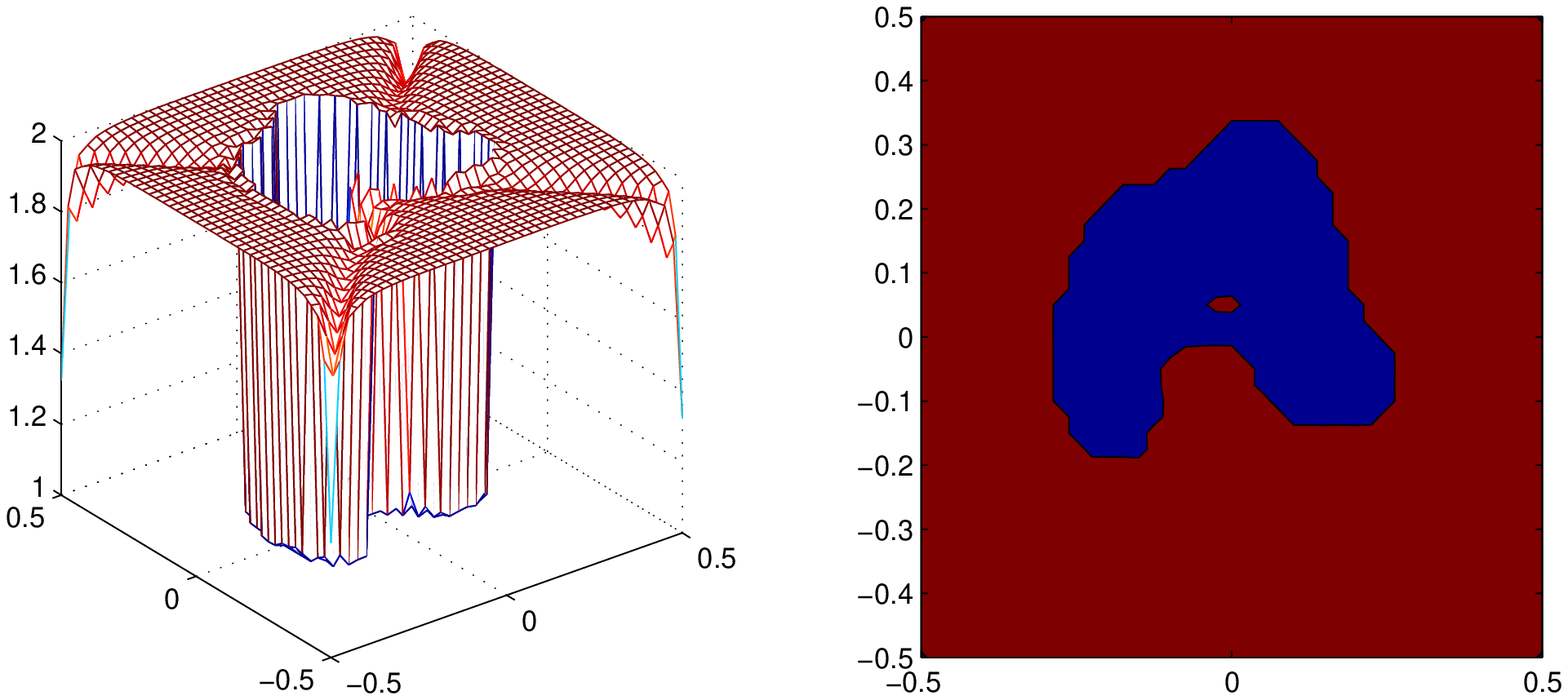}}
\hspace{0.3in} \subfigure[Step 150]{
\label{example2:e} %% label for fifth subfigure
\includegraphics[width=3.8in]{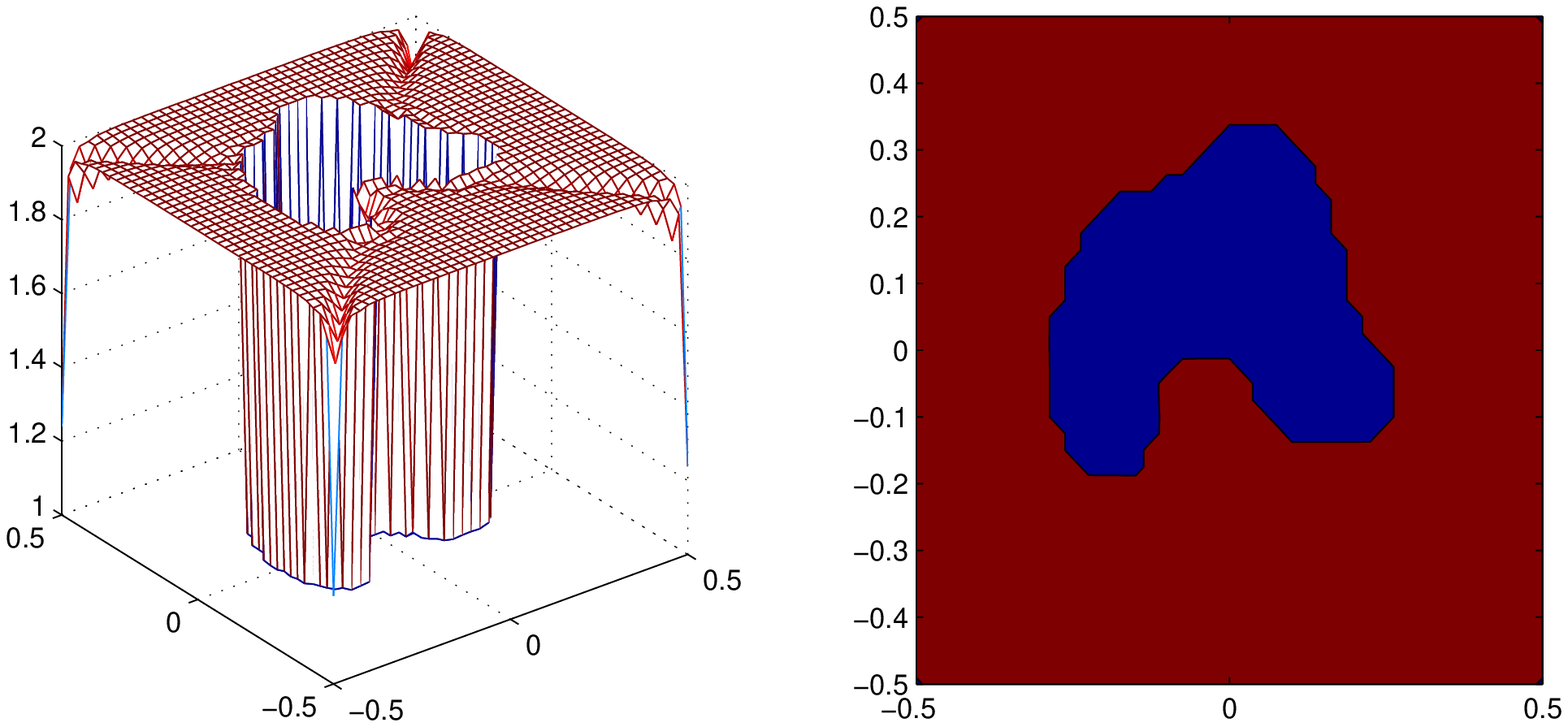}}
\hspace{0.3in} \subfigure[Step 200]{
\label{example2:f} %% label for sixth subfigure
\includegraphics[width=3.8in]{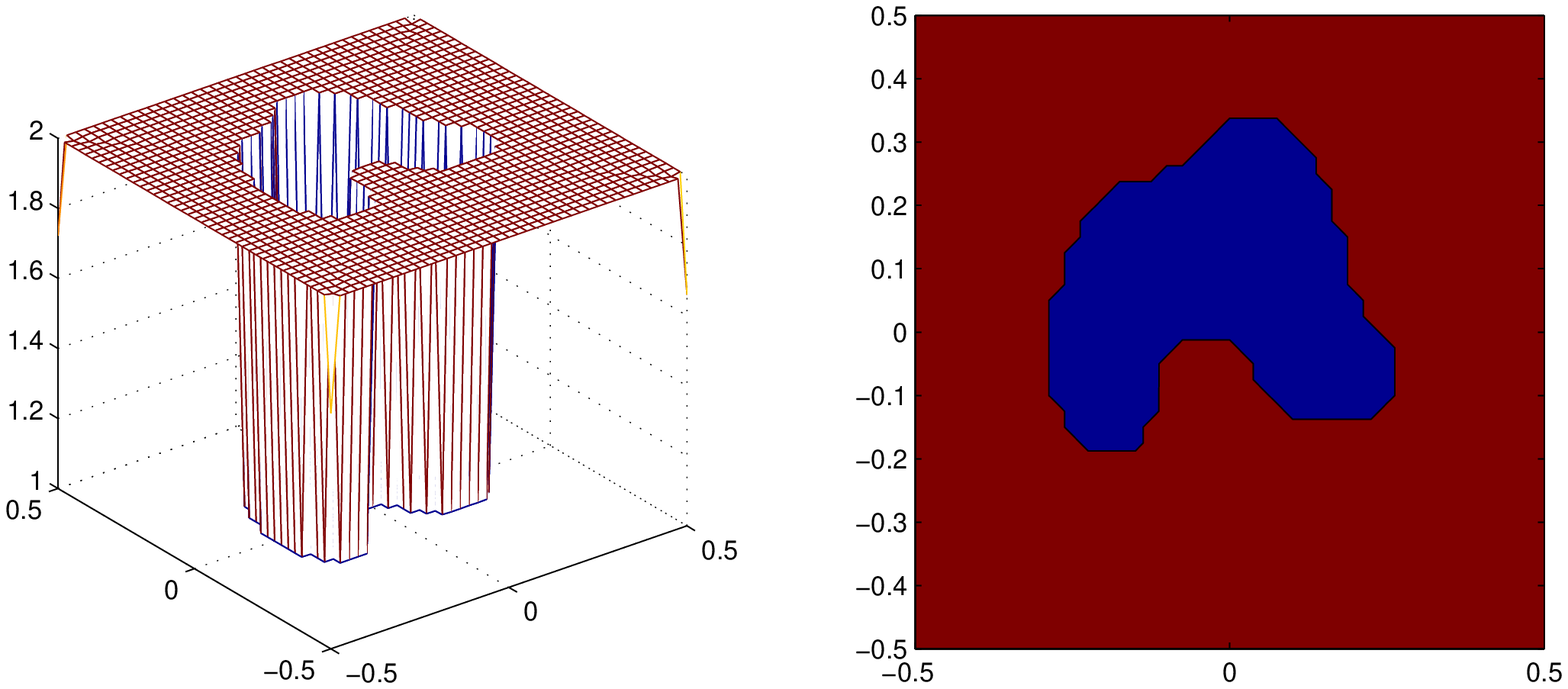}}
\caption{Numerical results for Example 2. (a) The final
reconstruction of $\phi$. (b) The change of the value of $F_1$ vs.
the number of iterations. (c)-(f) to show the evolution of $\phi$ and the interface.}
\label{example2:subfig} %% label for entire figure
\end{figure}

In this test, we set some variables as follows:
 $$dim=40,\quad
\alpha=0.001,\quad \sigma=0.9.$$
 The initial guess of $\phi$ and the upper bound of $osci$  are chosen as $\phi_{0}=1.5$ and $15$, respectively. The
 iterative process stops after 212 iterations. We present the numerical results
in Fig. \ref{example2:subfig}.

\begin{figure}
  \centering
  \includegraphics[width=3in]{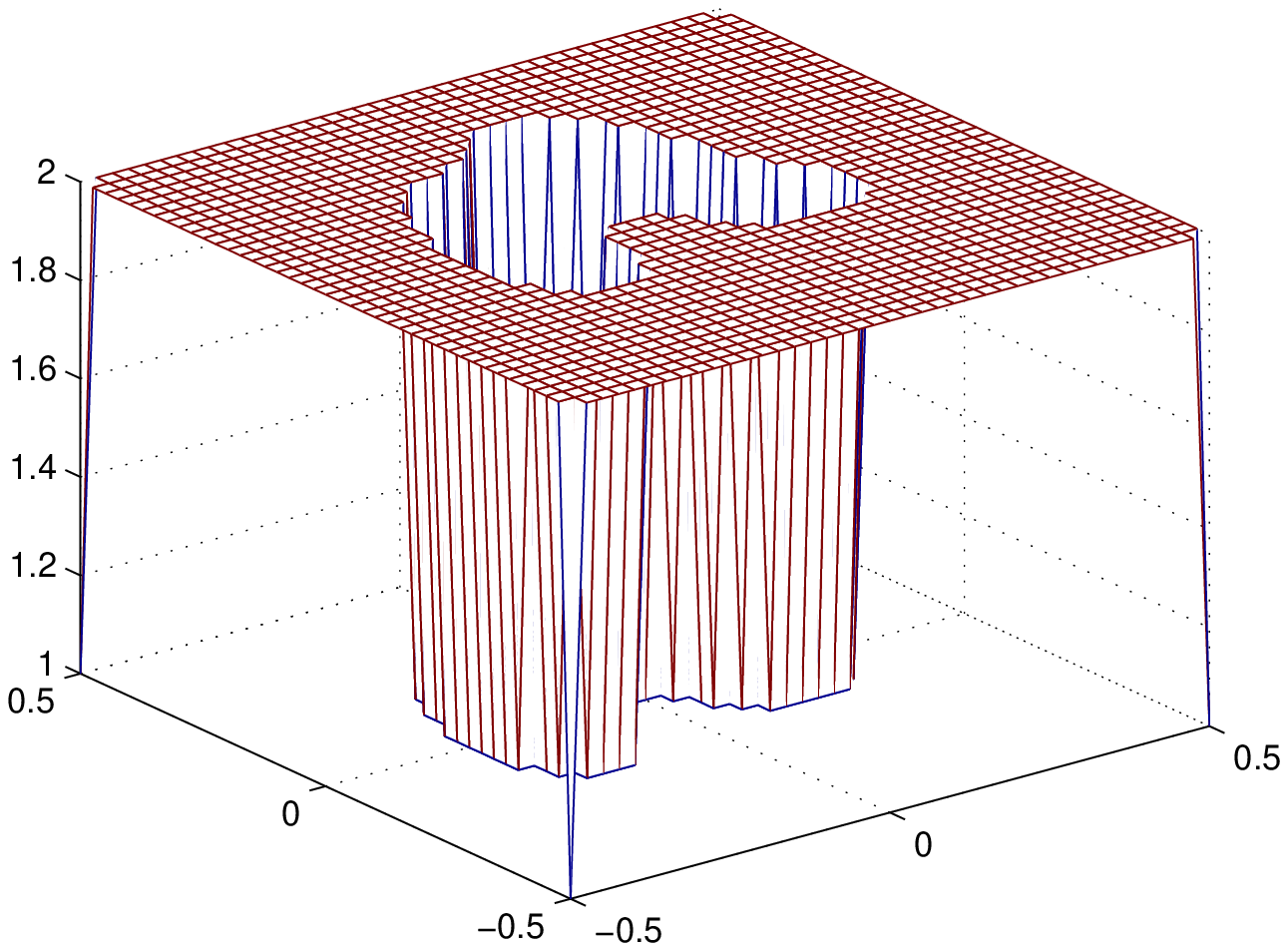}\\
  \caption{The figure of $\phi$ without regularization.}\label{example_2_without_reg}
\end{figure}

In Fig. \ref{example2:a}, the red solid line
coincides with the blue dotted one. And the number of the points
where $\phi$ and $\phi_{exact}$ take different values is 0. That is,  the algorithm can identify the shape of $D$ accurately. Fig.
 \ref{example2:b} depicts the change of $F_1$ with respect to the
number of iterations. In Fig. \ref{example2:b}, unlike Example \ref{example1}, the value of $F_1$ is below $10^{-11}$ after
some oscillations. Before
 exiting the iterative process, the times that the value of $F$ oscillates
do not exceed 15, so it's reasonable to choose the upper bound of
$osci$ to be 15. Fig. \ref{example2:c} to Fig. \ref{example2:f}
describe the evolution of $\phi$ and the interface. Similar to the evolution of the
level set function $\phi$ in the Example \ref{example1}, as the
increase of the number of iterations, the values of $\phi$ at four
corners first become smaller then start to increase to 2. In Fig.
\ref{example_2_without_reg}, the values of $\phi$ at four points of
$\Omega$ are 1. Thus we can conclude that it is the effects of the
regularization to pull the values of $\phi$ at these points back to
2.

%%%%%%%%%%%%%%%%%     Example-3      %%%%%%%%%%%%%%%%%
\begin{example}\label{example3}
In this example, in order to test the ability of identifying the
crack that is disconnected, we let the crack be the union of two circles and an ellipse. The current
density function  is chosen as $J=500$.
\end{example}

\begin{figure}
\centering \subfigure[The interfaces represented by $\phi_{exact}$ and $\phi$.]{
\label{example3:a} %% label for first subfigure
\includegraphics[width=2.7in]{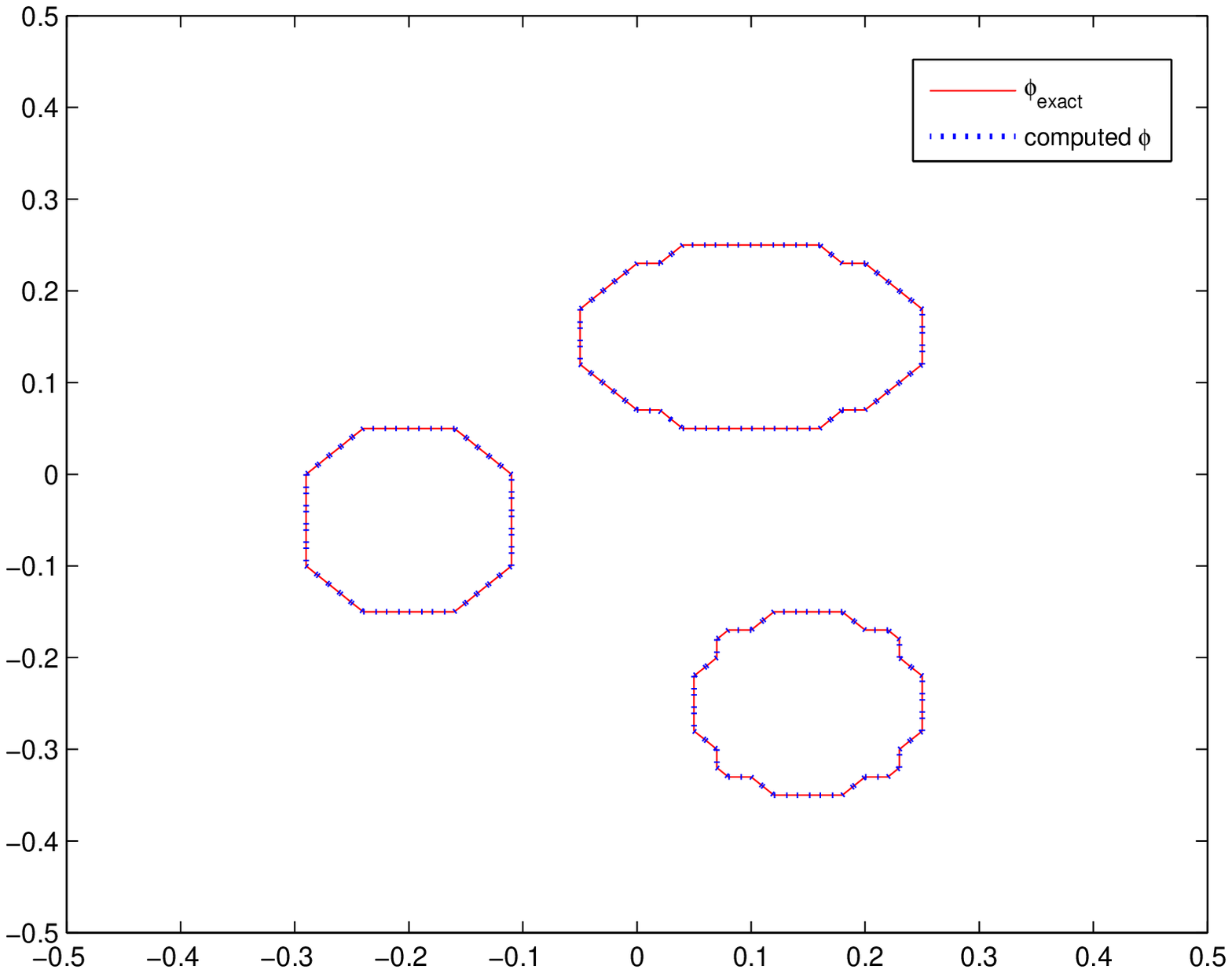}}
\hspace{0.1in} \subfigure[The value of $\log F_1/\log 10$.]{
\label{example3:b} %% label for second subfigure
\includegraphics[width=2.7in]{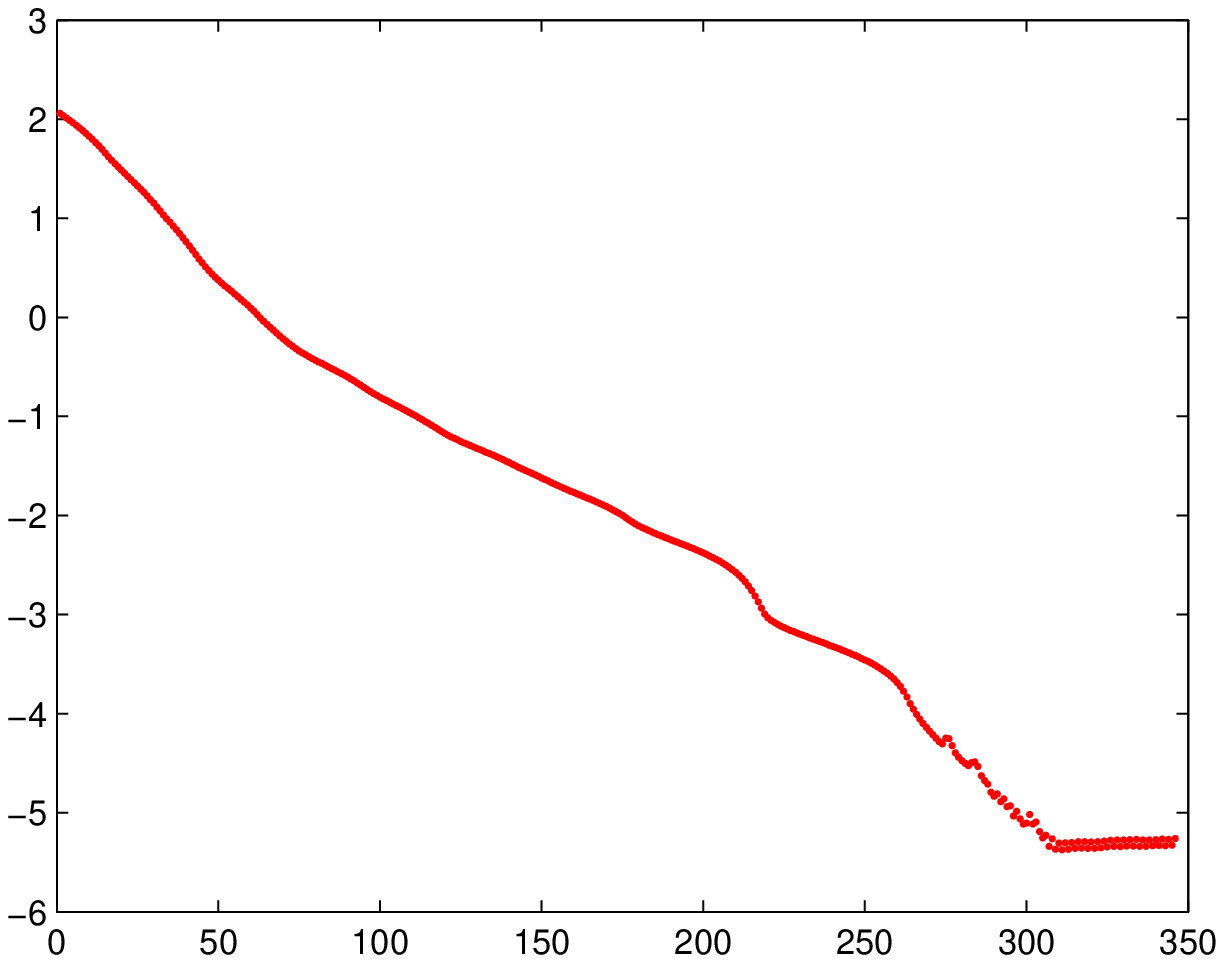}}
\hspace{0.1in} \subfigure[Step 75]{
\label{example3:c} %% label for second subfigure
\includegraphics[width=3.8in]{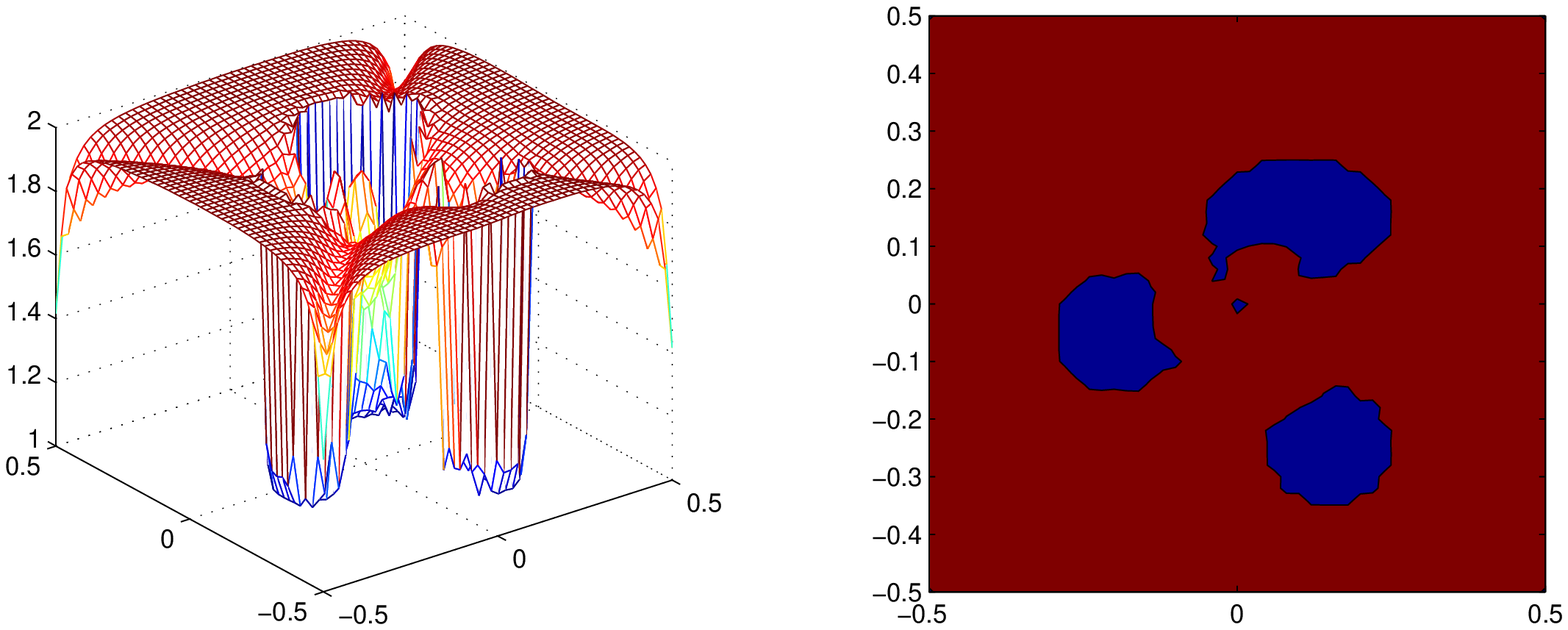}}
\hspace{0.1in} \subfigure[Step 150]{
\label{example3:d} %% label for second subfigure
\includegraphics[width=3.8in]{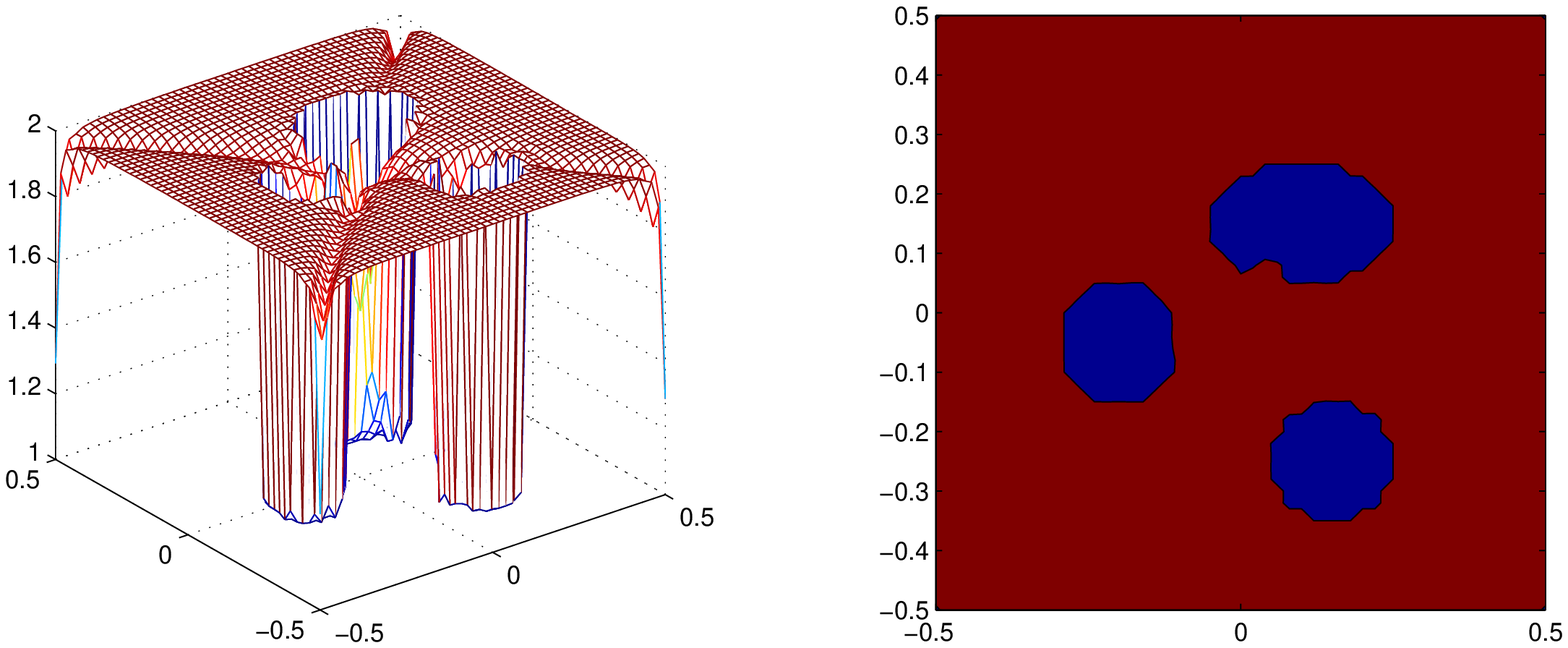}}
\hspace{0.1in} \subfigure[Step 225]{
\label{example3:e} %% label for second subfigure
\includegraphics[width=3.8in]{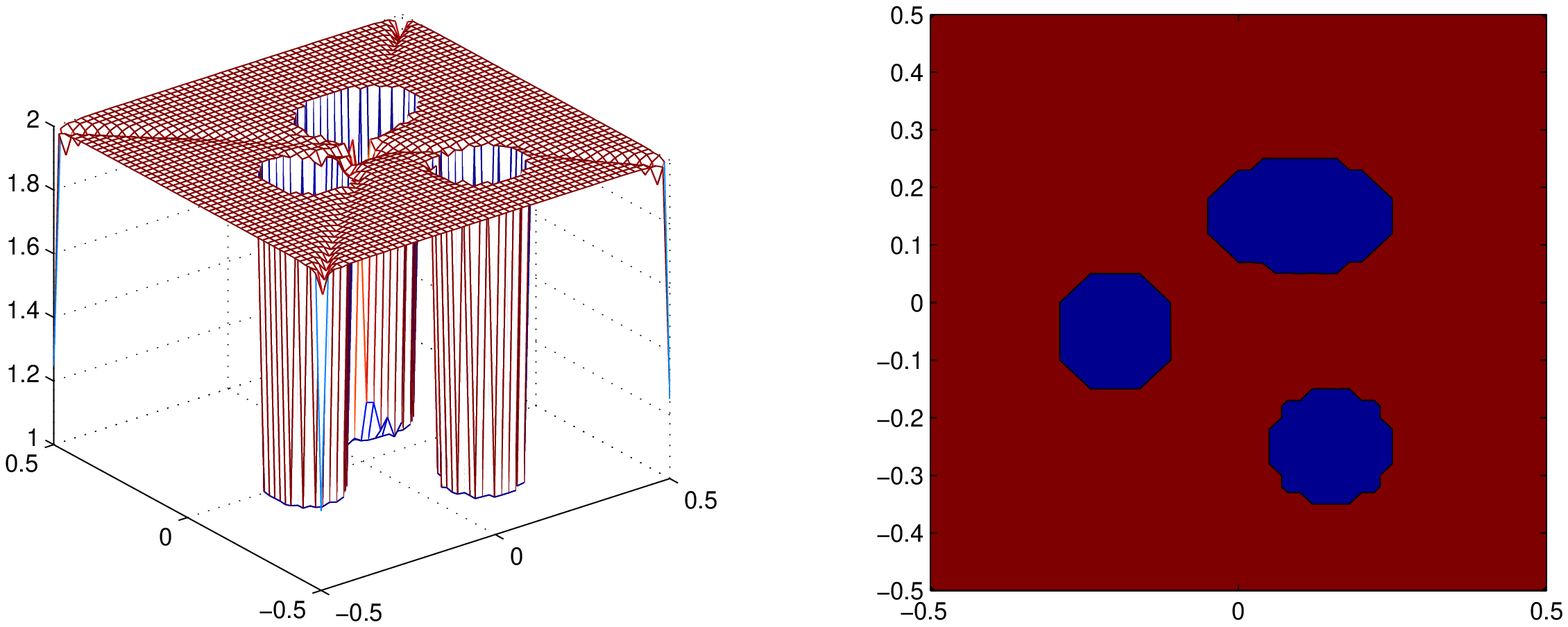}}
\hspace{0.1in} \subfigure[Step 300]{
\label{example3:f} %% label for second subfigure
\includegraphics[width=3.8in]{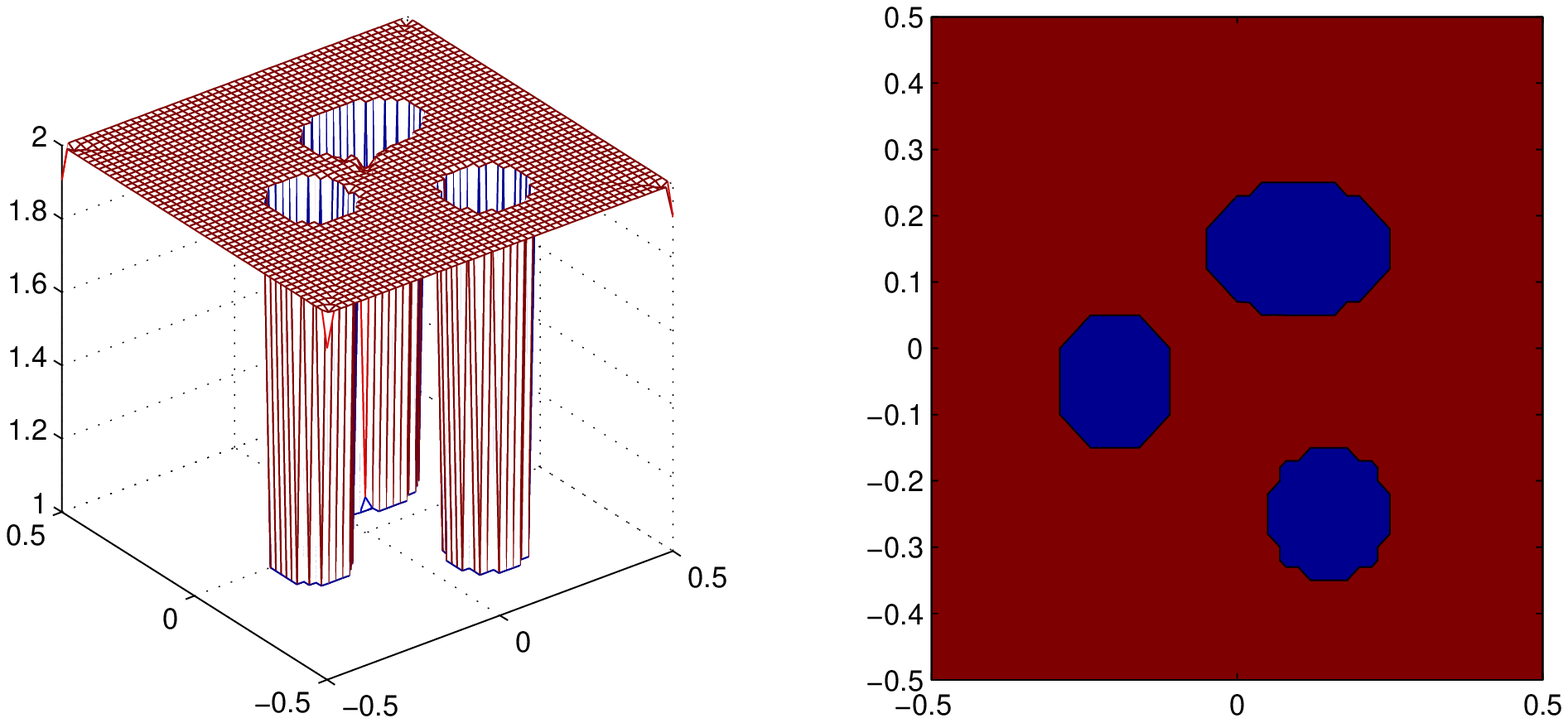}}
\caption{Numerical results for Example 3. (a) The final
reconstruction of $\phi$. (b) The change of the value of $F_1$ vs.
the number of iterations. (c)-(f) to show the evolution of $\phi$ and the interface.}
\label{example3:subfig} %% label for entire figure
\end{figure}

In this test, the variables are chosen as follows:
$$dim=50,\quad \alpha=0.001,\quad \sigma=0.9.$$
The initial value of the level set function and the upper bound of
$osci$ are $\phi_0=1.5$ and 10, separately. The iterative process
stops after 306 steps. We present the numerical results in Fig.
 \ref{example3:subfig}.

\begin{figure}
  \centering
  \includegraphics[width=3in]{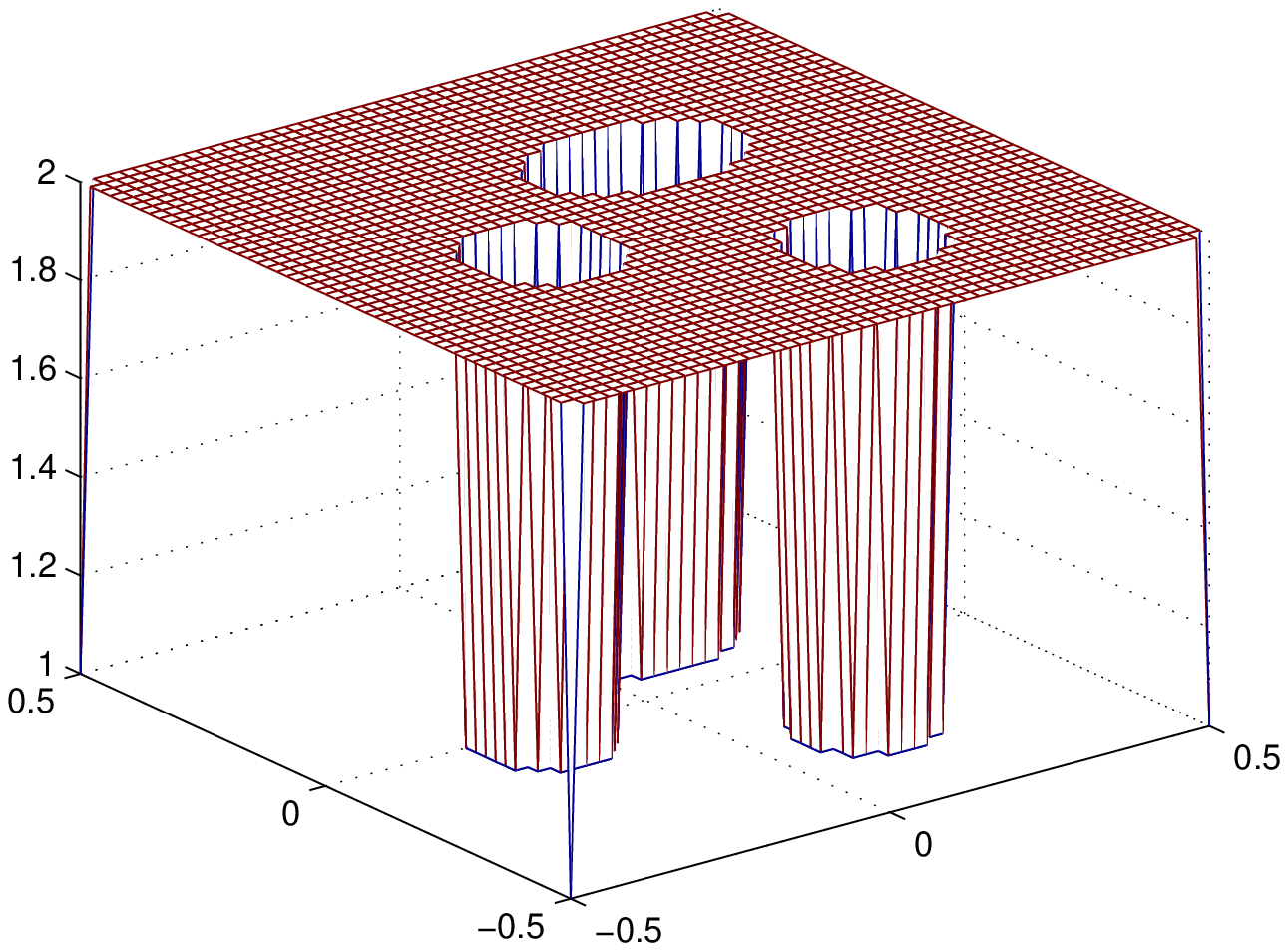}\\
  \caption{The figure of $\phi$ without regularization.}\label{example_3_without_regu}
\end{figure}

 In Fig. \ref{example3:a},  the interfaces between steel and air, represented by the red solid line and the blue dotted line, coincide. And the number of the points which
$\phi$ and $\phi_{exact}$ take different values is 0.  Fig.
\ref{example3:b} depicts the change of $F_1$ with respect to the
number of iterations. In Fig. \ref{example3:b},  the
value of $F_1$ first decreases and then oscillates in the last few steps. So
it is reasonable for us to choose 10 as the upper bound of $osci$.
Fig. \ref{example3:c} to Fig. \ref{example3:f} describe the
evolution of $\phi$ and the interface. As the increase of the number of iterations,
the values of $\phi$ at four corners first become smaller then start
to increase to 2. In Fig. \ref{example_3_without_regu}, the values
of $\phi$ at four points of $\Omega$ are 1. Thus we can conclude
that it is the effects of the regularization term to pull the values
of $\phi$ at these points to 2.

Compared with the algorithm in \cite{IC_ref}, PCLSA takes much less steps, but reconstructs a more accurate shape. In \cite{IC_ref}, it takes 574 iteration steps to stop the algorithm. In our PCLSA, we only need half of that iterations, that is, totally 307 iterations to stop the algorithm. Moreover, with more iterations in \cite{IC_ref} the final reconstruction shape still deviates a lot from the exact shape. In our final result (see Fig \ref{example3:subfig}), however, there is no deviations at all. Thus PCLSA performs better than the algorithm in \cite{IC_ref} for this example.

\begin{figure}
\centering \subfigure[The final result for $\phi_0=1.2$.]{
\label{flexi_ini:a} %% label for first subfigure
\includegraphics[width=2.7in]{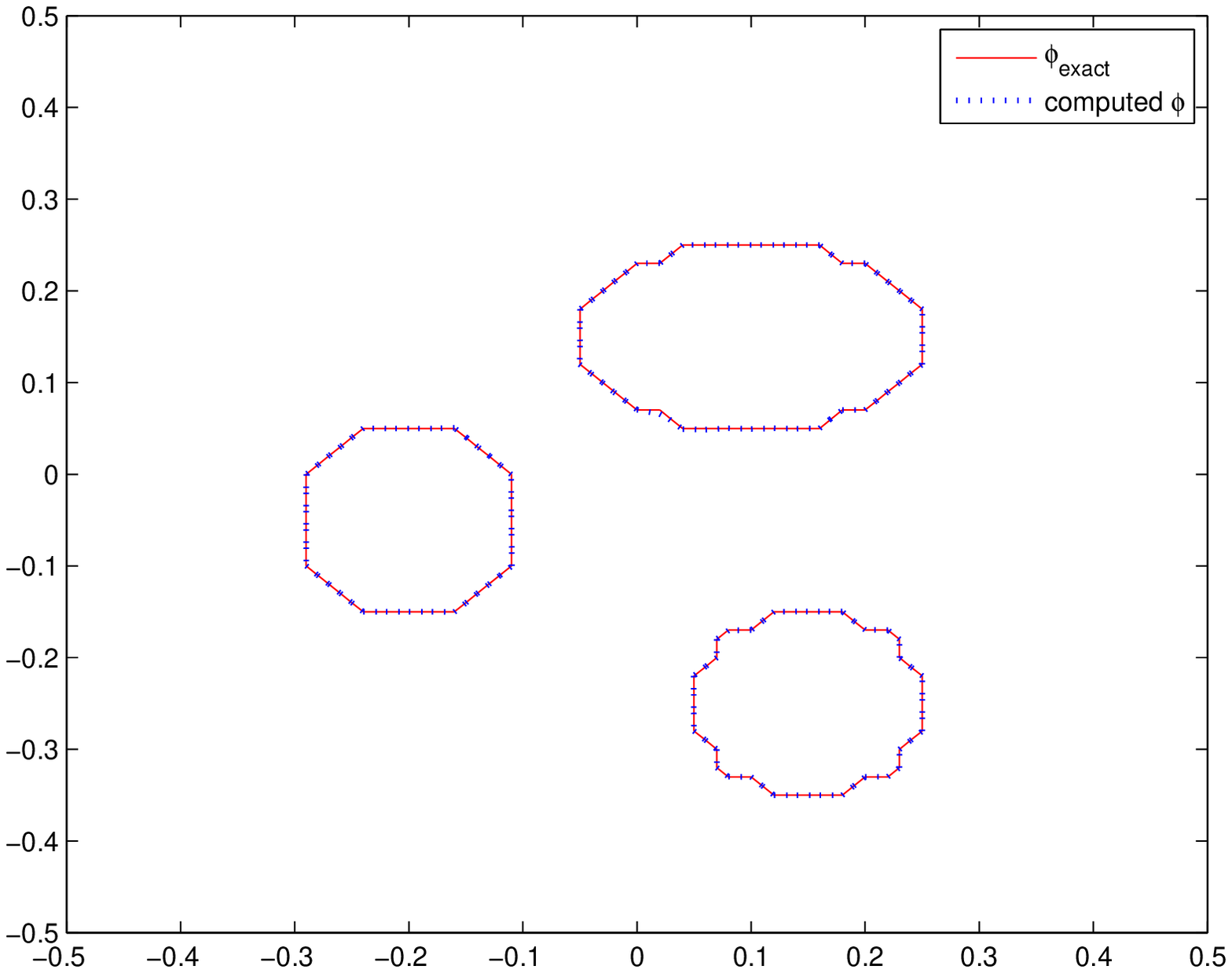}}
\hspace{0.1in} \subfigure[The final result for $\phi_0=1.4$.]{
\label{flexi_ini:b} %% label for second subfigure
\includegraphics[width=2.7in]{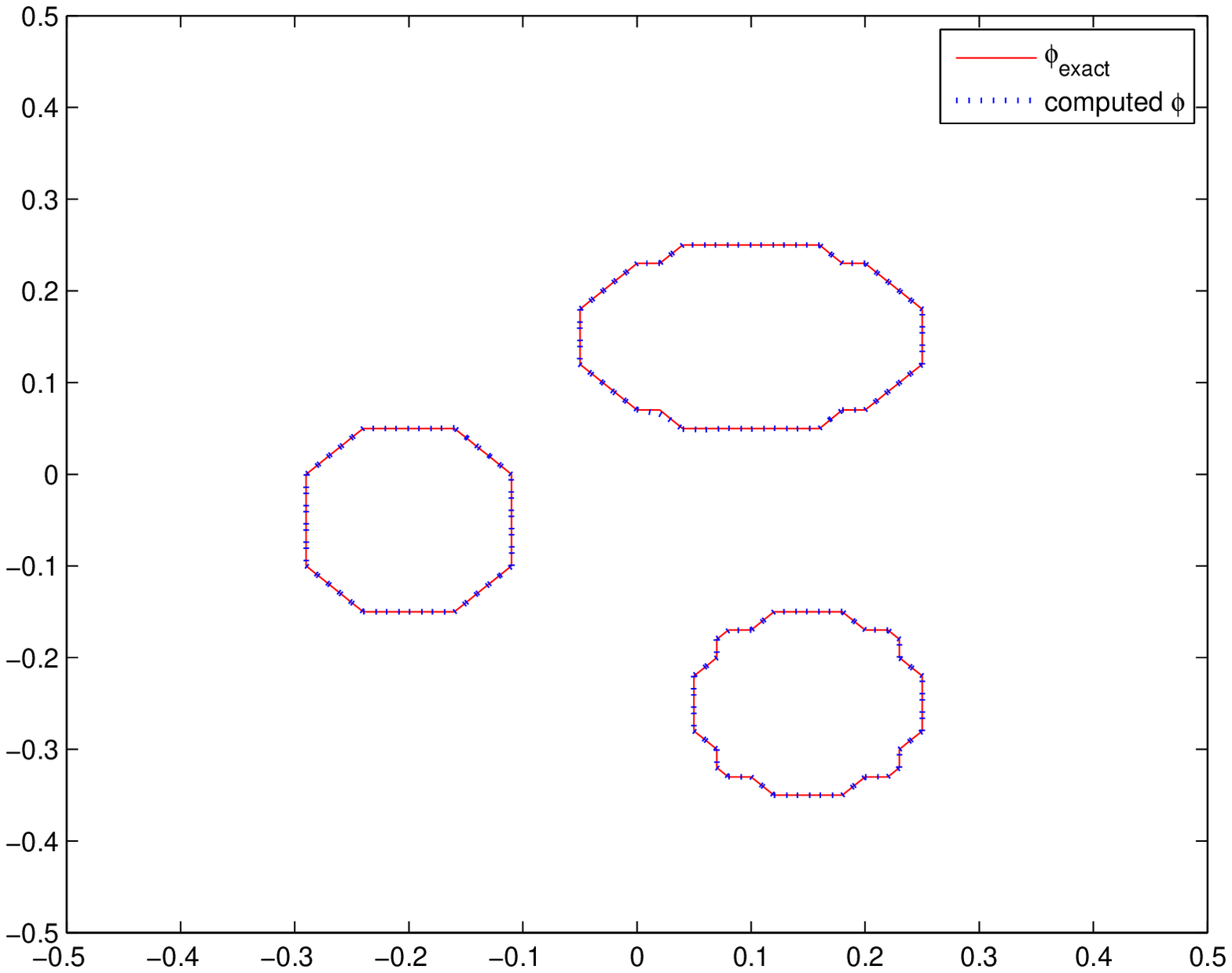}}
\hspace{0.1in} \subfigure[The final result for  $\phi_0=1.7$.]{
\label{flexi_ini:c} %% label for second subfigure
\includegraphics[width=2.7in]{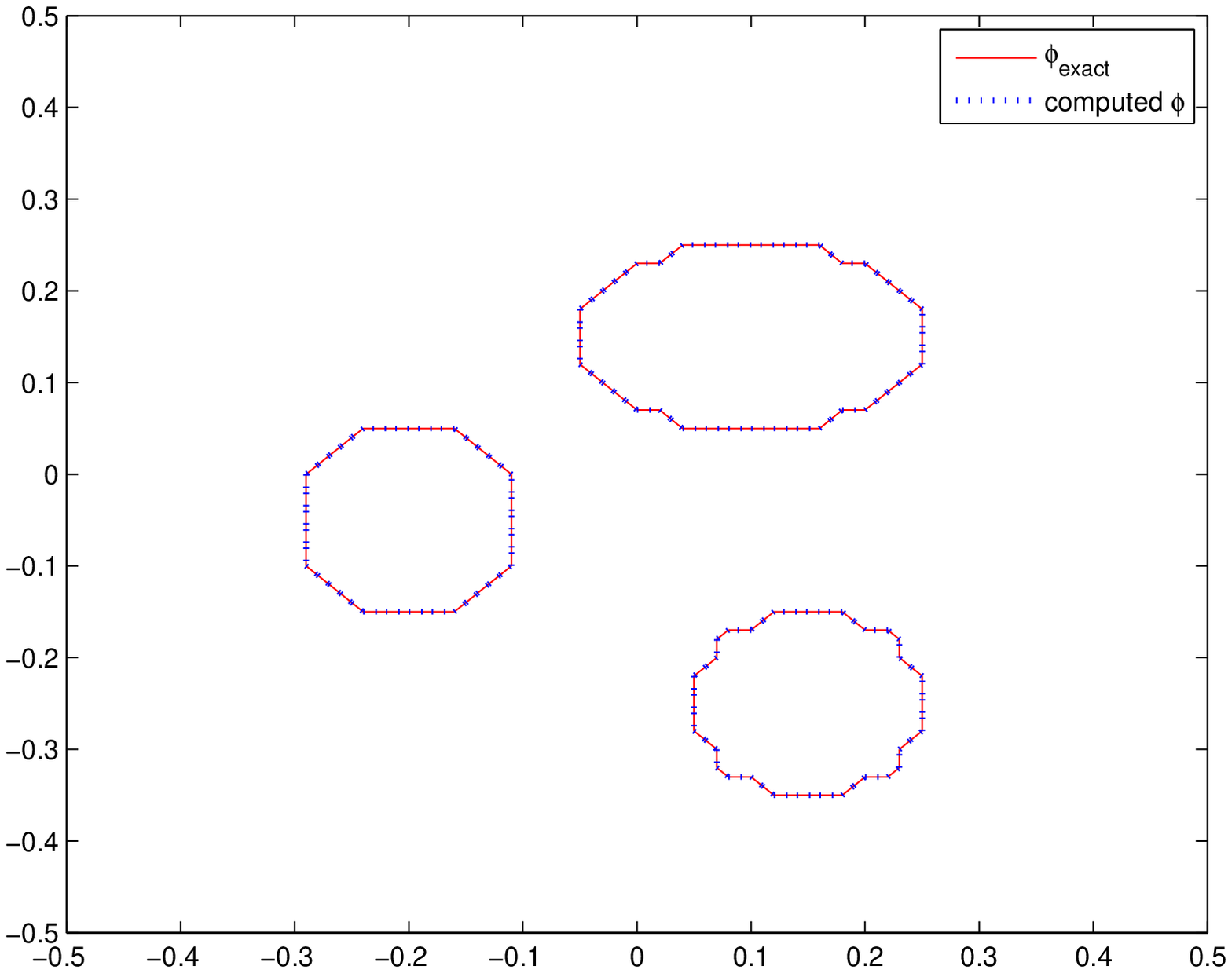}}
\hspace{0.1in} \subfigure[The final result for  $\phi_0=1.9$.]{
\label{flexi_ini:d} %% label for second subfigure
\includegraphics[width=2.7in]{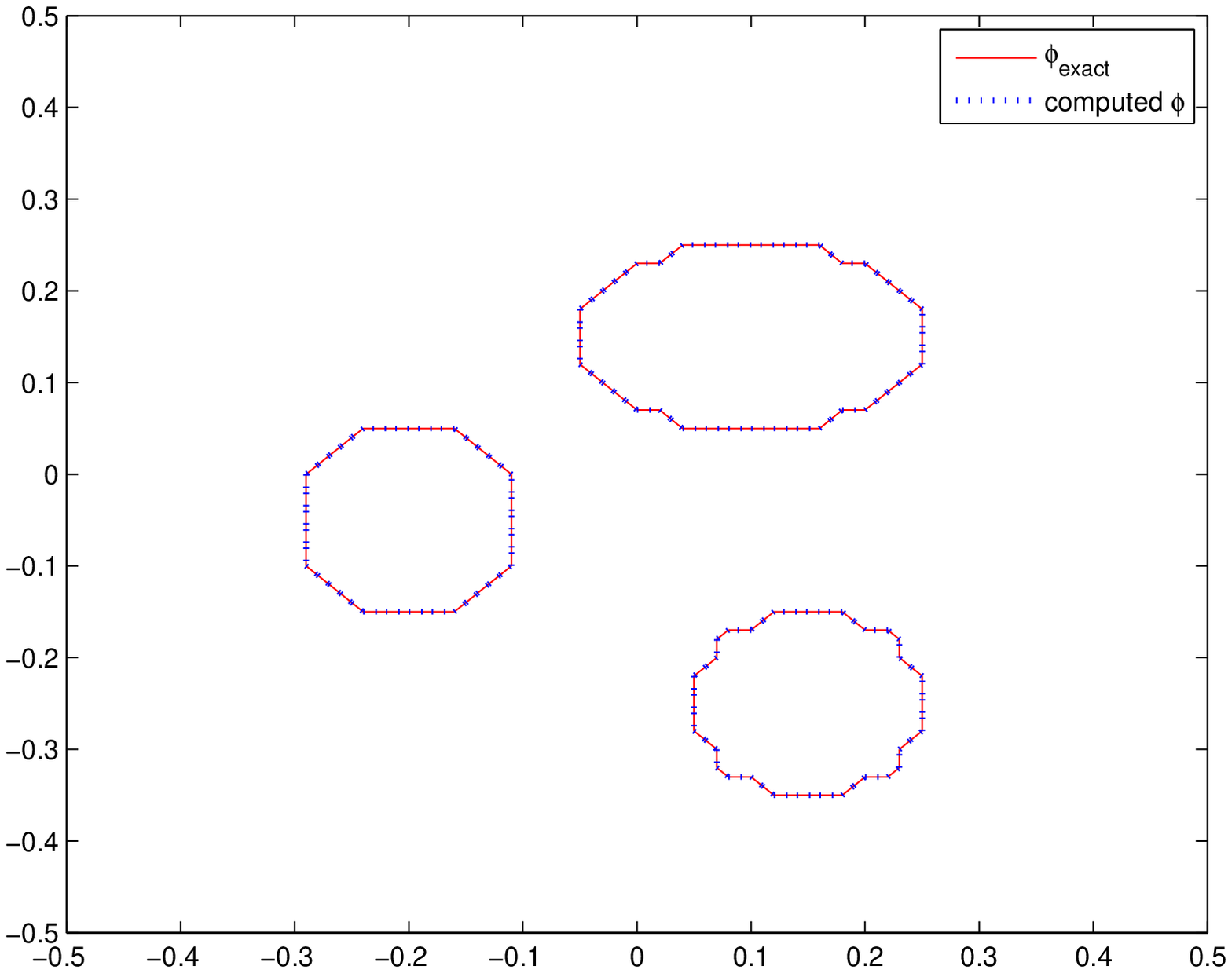}}
\caption{The final recovery interfaces of different values of $\phi_0$.}
\label{flexi_ini:subfig} %% label for entire figure
\end{figure}

\begin{figure}
\centering \subfigure[The final result for for Case 1.]{
\label{flexi_rand:a} %% label for first subfigure
\includegraphics[width=2.7in]{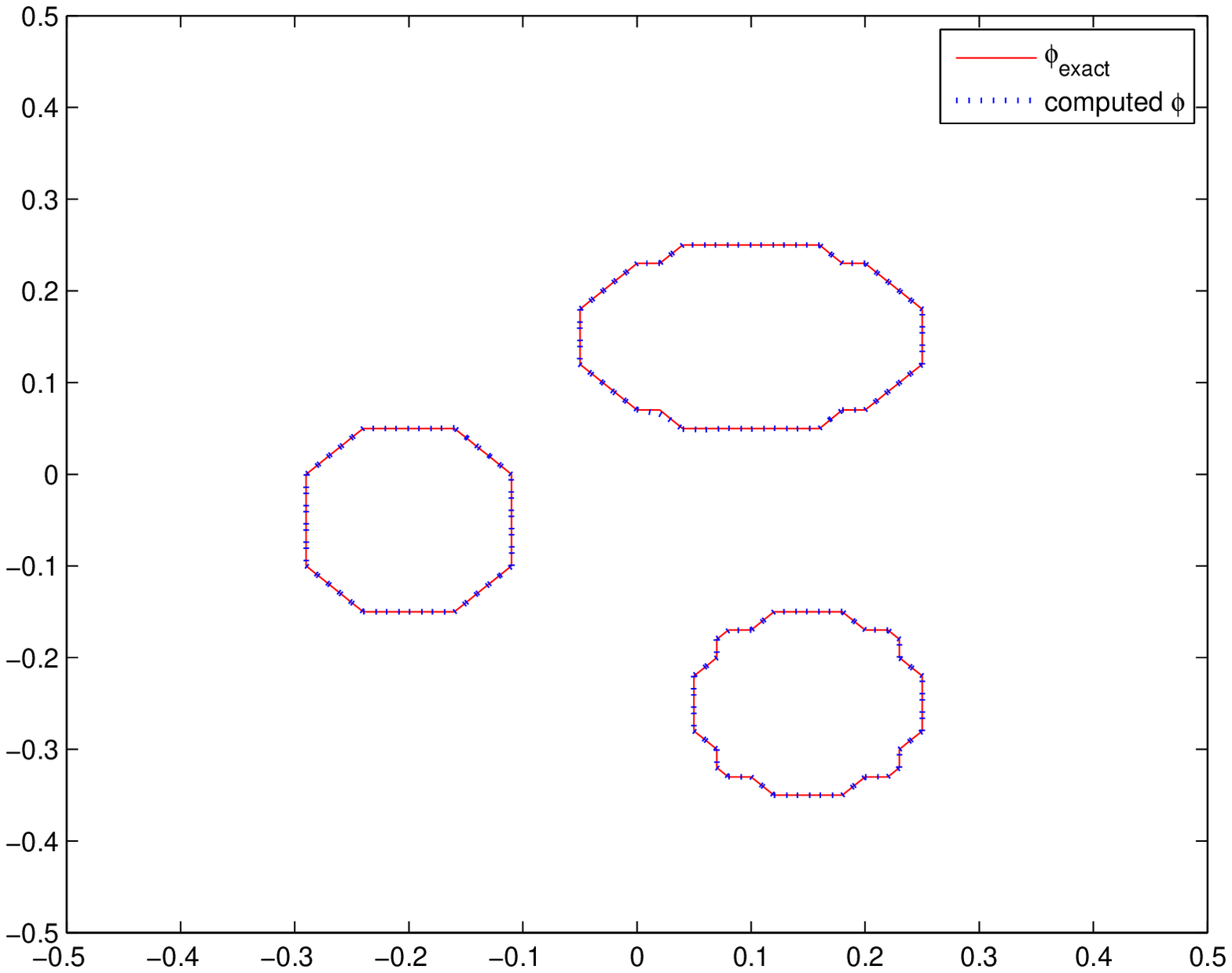}}
\hspace{0.1in} \subfigure[The final result for Case 2.]{
\label{flexi_rand:b} %% label for second subfigure
\includegraphics[width=2.7in]{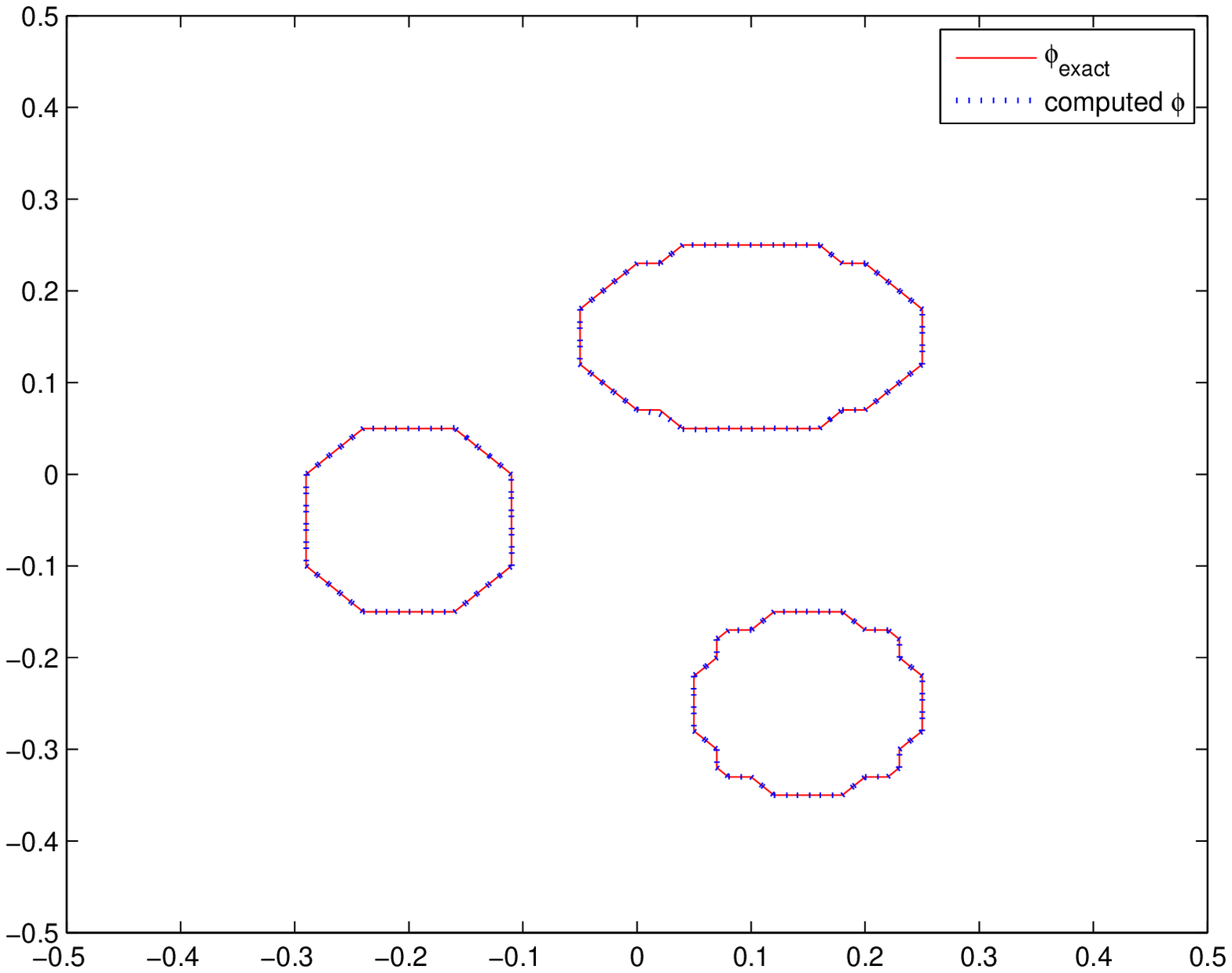}}
\caption{The final recovery interfaces of two cases of rand initial guesses: $\phi_0(x)=1+rand(x)$.}
\label{flexi_rand:subfig} %% label for entire figure
\end{figure}

 In order to show the flexibility of the initial guess of $\phi$, we assign different values to $\phi_0$ and compare the final results. we set $\phi_0$  to be different constants, that is,  1.2, 1.4, 1.7, 1.9,
respectively, and present the final results in Fig. \ref{flexi_ini:subfig}. For these four different values of $\phi_0$, the interfaces represented by $\phi$ and $\phi_{exact}$
coincide and the number of the points where $\phi$ and $\phi_{exact}$ take different value is 0. This indicates that our algorithm can
identify the shape exactly. Furthermore, we choose a more general function as the initial value of $\phi$ and also observe the final interface. The general function is $\phi(x) = 1+rand(x)$ where $rand(x)$ can produce pseudo-random values between 0
and 1. We present the numerical results in Fig. \ref{flexi_rand:subfig}, where the interfaces represented by $\phi$ and $\phi_{exact}$ coincide.

From the above observation, PCLSA relies little on the initial value of the level set function.
Thus, we  need not to put too much attentions on the initial guess of $\phi$ in our PCLSA for the nonlinear electromagnetism recovery problems.

\begin{example}\label{example4}
To test the robustness of PCLSA, we use the measurement data
of $\overline{\mathbf{M}}$ with  a certain level of noise.
For this test, we use the same case in example 2 except that the
measurement data of $\overline{\mathbf{M}}$ is polluted by the
certain level of noise, 5\%, 10\%,15\% and 20\%.
\end{example}

\begin{figure}
\centering \subfigure[The final result for $5\%$ noise]{
\label{example4:a} %% label for first subfigure
\includegraphics[width=2.7in]{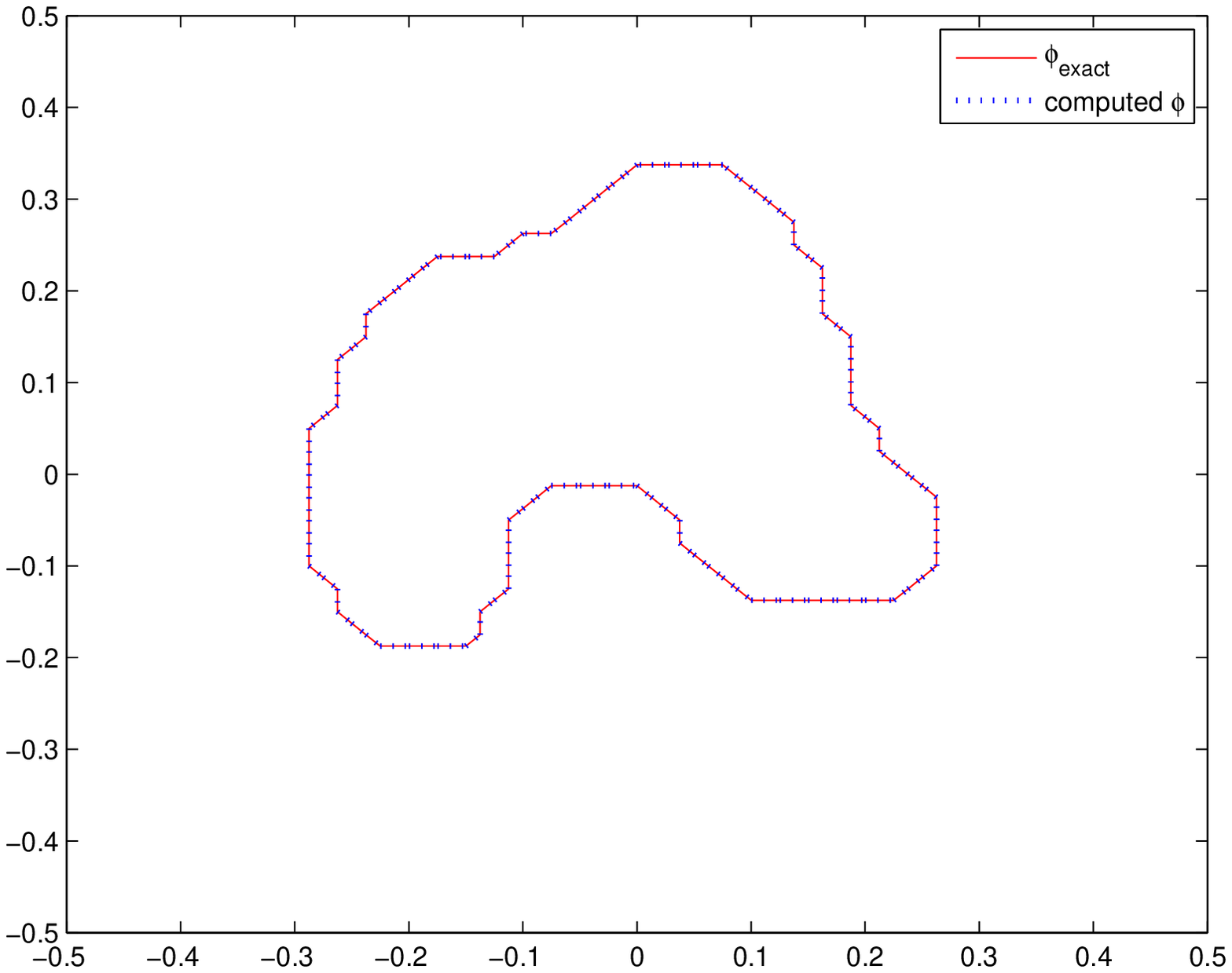}}
\hspace{0.1in} \subfigure[The final result for $10\%$ noise.]{
\label{example4:b} %% label for second subfigure
\includegraphics[width=2.7in]{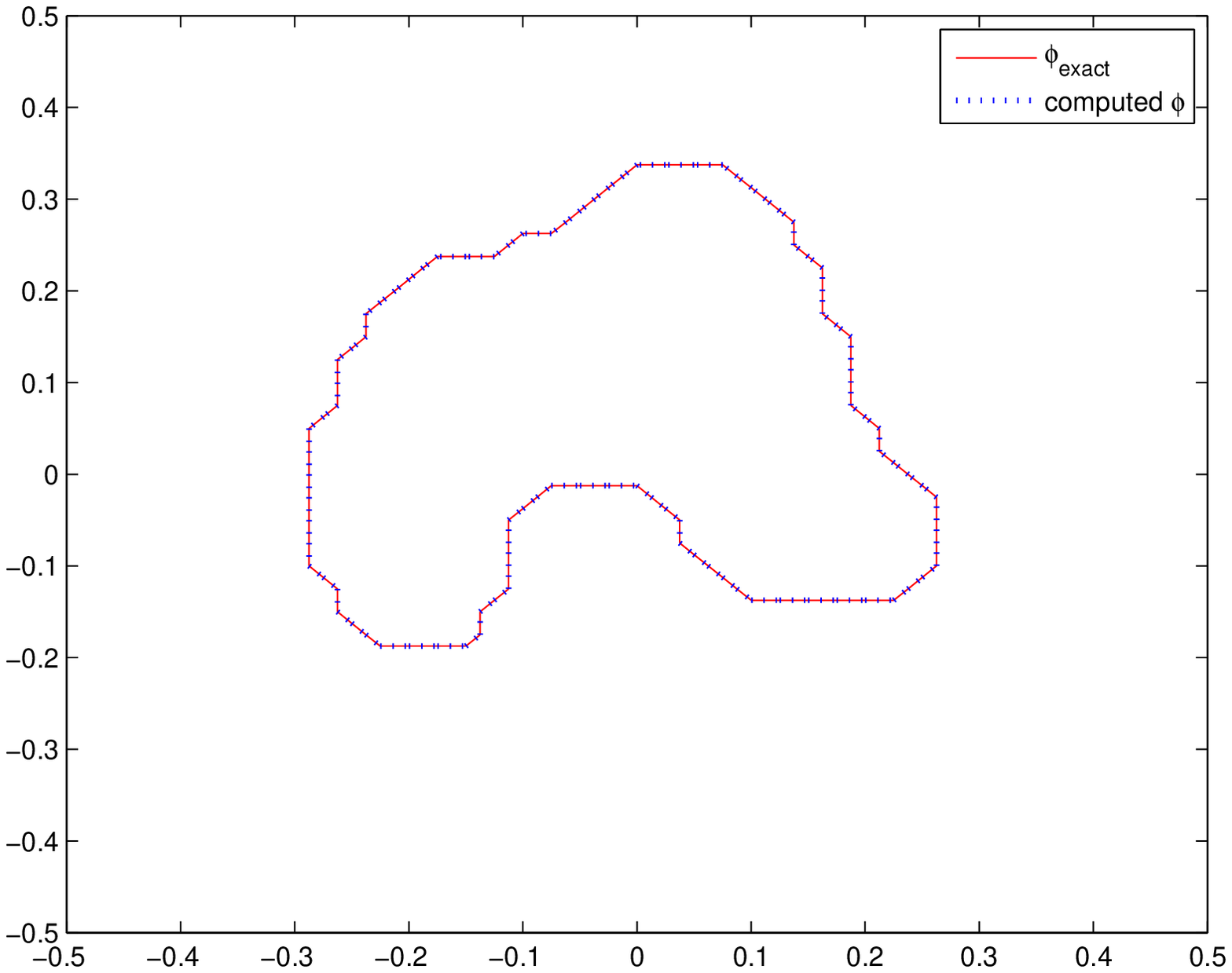}}
\hspace{0.1in} \subfigure[The final result for $15\%$ noise.]{
\label{example4:c} %% label for second subfigure
\includegraphics[width=2.7in]{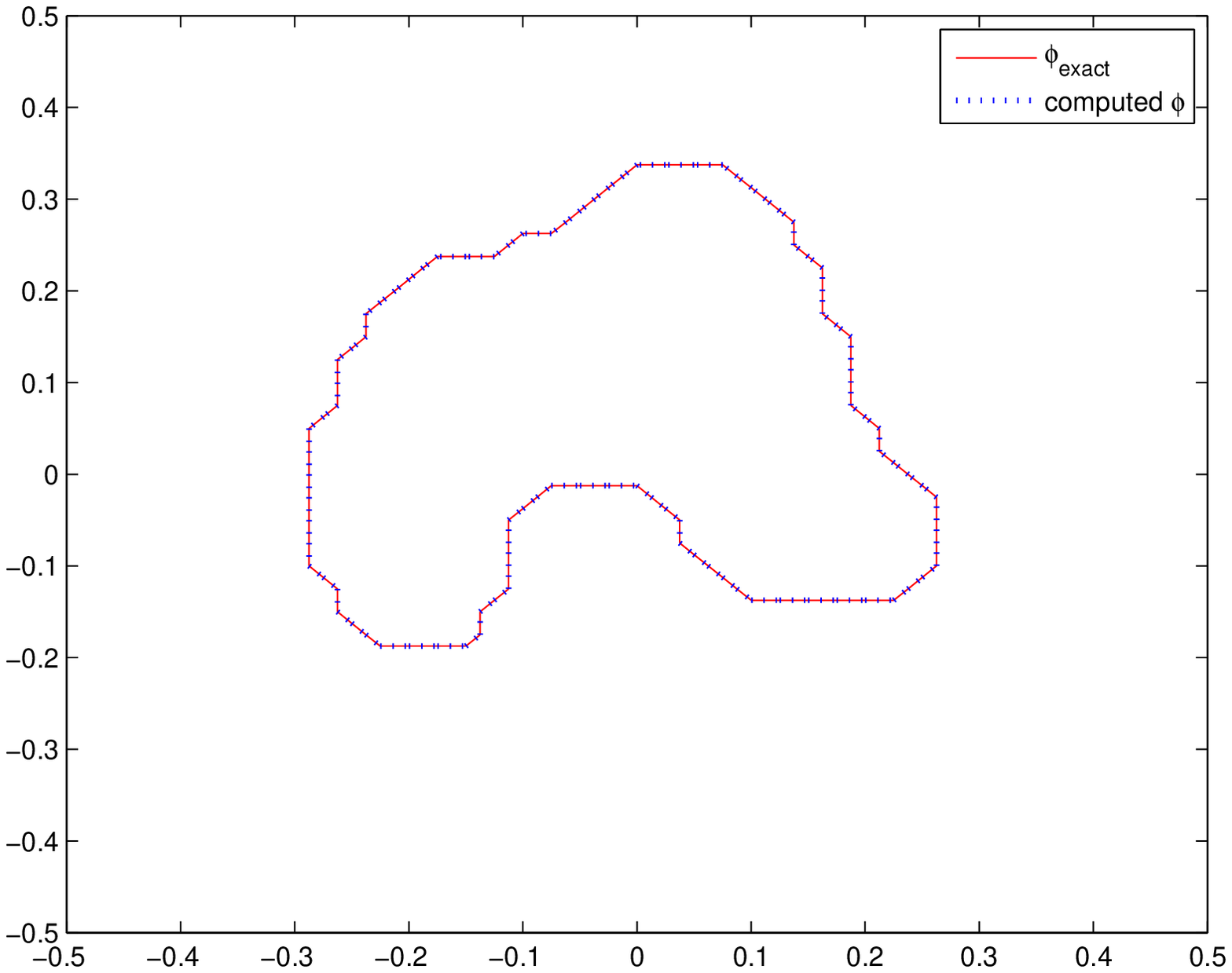}}
\hspace{0.1in} \subfigure[The final result for $20\%$ noise.]{
\label{example4:d} %% label for second subfigure
\includegraphics[width=2.7in]{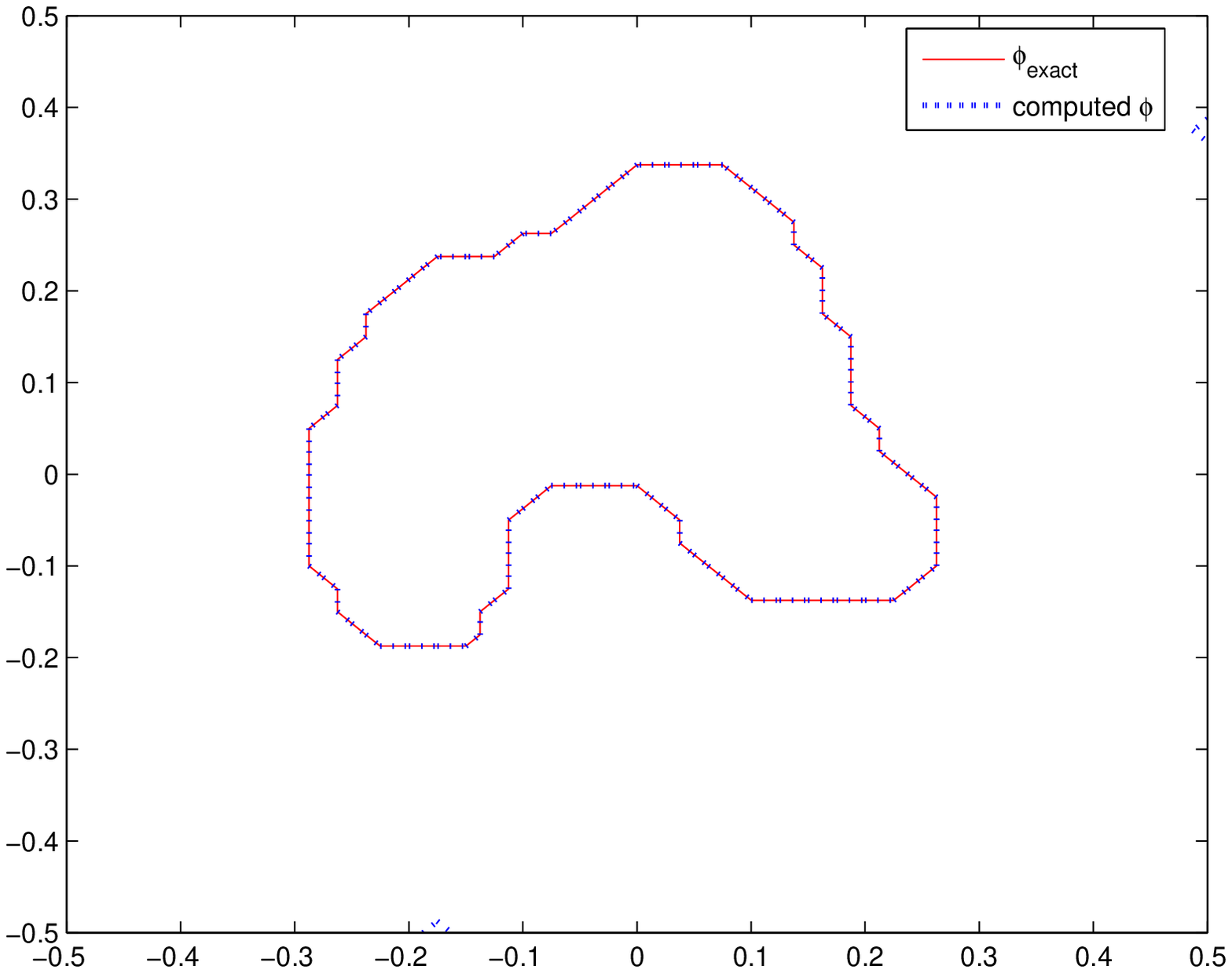}}
\caption{ The final recovery interfaces of different noise levels in Example 4.}
\label{example4:subfig} %% label for entire figure
\end{figure}

In this test,  $dim=50$. We set $\phi_0=1.5$, the upper bound of $osci$  to be 10, $\alpha=0.1$ and
$\sigma=0.9$. For  different noise level cases,  the final results are shown in Fig. \ref{example4:subfig}.
When the noise level is under 15\%, the PCLSA reconstructs the shape of
$D$ completely. That is, in our numerical test,  the two functions $\phi$ and $\phi_{exact}$ take the
same value at every point $\emph{\textbf{x}}\in\Omega$. But when the
noise level is up to 20\%, though the interface between the steel and air
represented by $\phi$ is recovered, there are some flaws at the
interface between $D$ and $\Omega\backslash D$ (some blue points). For the range of $\sigma$,
the interval $0.6\sim0.9$ is shown by our numerical tests to be
not a bad choice.

% We will next investigate the effects of the constant $C_1$ in the
%initial value of $\phi$ on the final result to show the flexibility
%of the initial value of $\phi$. For Example 3, we use the setting in
%Example 3 but with the constant in the initial value of $\phi$ being
%1.2, 1.4, 1,7, 1.9, respectively. We present the numerical result in
%Fig. \ref{flexi:subfig}. For every choice of the initial value of
%$\phi$, the interface represented by $\phi$ and $\phi_{exact}$
%coincide and the number of the points where $\phi$ and
%$\phi_{exact}$ take different value is 0. This indicates that our
%algorithm can identify the shape exactly.

However, $\sigma$ affects the iteration steps in our PCLSA. The larger
$\sigma$ is, the less steps  the algorithm takes. But at high noise level, the reconstructed shape is better when the value of $\sigma$ is smaller.
%In example 4,
%our PCLSA  find the shape inaccurately with some flaws at
%the boundary when $\sigma=0.9$ and the noise
%level is 15\%, but can find the shape accurately when $\sigma=0.7$.

\section{Conclusion}
In this paper, we propose a piecewise constant level set algorithm for the nonlinear electromagnetism inverse problem.
In our numeric test, our PCLSA do not rely on  the initial guess of the level set function $\phi$, and
can reconstruct the shape of $\Omega$ exactly even when the noise
level is high. Our PCSLA is quite effective and robust to solve a kind of nonlinear inverse problems.
%%%%%%%%%%%%%%%%%%%%%%%%%%%%%%% New Section   %%%%%%%%%%%%%%%%%%%%%%%%%%%%%%%%%%%%%%%%%%%%%%%%%%%%%%%%%%%%%%

%%%% Acknowledgments %%%%%%%%
\section*{Acknowledgments}
We express our  gratitude to Doctor Ivan Cimr\^{a}k for his prompt and  comprehensive  answer to our problems. This work is supported by Natural Science Foundation of Zhejiang Province, China (No. LY12A01023), National Natural Science Foundation of China (No. 11201106) and Natural Science Foundation of Zhejiang Province, China (No. LQ12A01001).

%%%% Bibliography  %%%%%%%%%%

\end{document}